\documentclass{article}
\usepackage[a4paper, portrait, top=2cm,bottom=2cm,left=3cm,right=3cm,marginparwidth=1.75cm]{geometry}
\usepackage[english]{babel}
\usepackage[utf8]{inputenc}
\usepackage{amsfonts}
\usepackage{amsmath}
\usepackage{amsthm}
\usepackage{amssymb}
\usepackage{bbm}
\usepackage{bm} 
\usepackage{booktabs}
\usepackage[font={small,it}]{caption} 
\usepackage[autostyle,italian=quotes]{csquotes} 
\usepackage{empheq}
\usepackage{emptypage}
\usepackage{float}
\usepackage{graphicx}
\usepackage{indentfirst}
\usepackage{latexsym}
\usepackage{lettrine}
\usepackage{mathtools} 
\usepackage{microtype} 
\usepackage{braket}
\usepackage{relsize}
\usepackage{siunitx} 
\usepackage{subcaption}
\usepackage{tikz} 
\usetikzlibrary{arrows.meta}
\usepackage{chemfig}
\usepackage{tocloft}
\usepackage{notoccite}
\usepackage{geometry}
\usepackage{fancyhdr} 
\usepackage[version=4]{mhchem}
\usepackage{hyperref}
\usepackage{comment}

\DeclareMathOperator{\sgn}{sgn}
\DeclareSIUnit{\molar}{M}

\newcommand{\vb}[1]{\bm{#1}}  

\fancypagestyle{plain}{
	\fancyhf{}

}

\makeatletter
\patchcmd{\@maketitle}{\LARGE \@title}{\fontsize{14}{19.2}\selectfont\@title}{}{}
\makeatother

\usepackage{authblk}

\newsavebox\affbox
\author[1]{\textbf{Valentina Buonfiglio}}
\author[2]{\textbf{Irene Pertici}}
\author[2]{\textbf{Pasquale Bianco}}
\author[3]{\textbf{Duccio Fanelli}}
\author[1]{\textbf{Stefano Gherardini}}
\affil[1]{Istituto Nazionale di Ottica, Consiglio Nazionale delle Ricerche (CNR-INO), Largo Enrico Fermi 6, 50125, Firenze (FI), Italy}
\affil[2]{PhysioLab, University of Florence, Sesto Fiorentino (FI), Italy}
\affil[3]{Department of Physics and Astronomy, University of Florence, Sesto Fiorentino (FI), Italy}

\title{Reverse engineering of mechano-kinetic parameters from stochastic force profiles in heterogeneous ensembles of molecular motors and tracks}

\date{}    

\begin{document}
    
\pagestyle{headings}	
\newpage
\setcounter{page}{1}
\renewcommand{\thepage}{\arabic{page}}
	
\maketitle
	
\noindent\rule{15cm}{0.5pt}
    \begin{abstract}
    Heterogeneity in contractile systems of molecular motors interacting with their tracks plays a key role in numerous cellular physiological and pathological processes. However, its effects cannot be captured by theoretical models assuming identical mechano-kinetic properties for all motors. We developed a stochastic framework to describe heterogeneous ensembles of myosin motors comprising two populations with distinct mechano-kinetic properties. Assuming that motors interact with the actin filament (their track) as independent force generators, we derived the probability distribution of the isometric force and characterised the statistics of finite-size force fluctuations. The proposed framework includes an estimation procedure that simultaneously infers the mechano-kinetic parameters and the size of the motor ensemble, eliminating the need to prescribe it a priori. Validation against synthetic and experimental data shows the model accurately captures force fluctuations and provides realistic estimates of the ensemble size. Furthermore, the framework enables a quantitative assessment of ensemble heterogeneity and the inference of an unknown motor species properties. This approach provides a quantitative tool for characterising heterogeneous actin-myosin systems, such as those arising from the co-presence of different protein isoforms in cardiac and skeletal muscle, or from the partial replacement of native proteins with mutation-derived or engineered variants.

    \let\thefootnote\relax\footnotetext{
    \small $^{*}$\textbf{Corresponding author} 
    \textit{\color{cyan} valentina.buonfiglio@unifi.it}}
    \vspace{0.8 cm}
    \end{abstract}
    
\noindent\rule{15cm}{0.4pt}

\section{Introduction}

Molecular motors convert biochemical energy released by ATP hydrolysis into work.
Among them, myosin II is the molecular motor that, interacting in ensemble with its track, the actin filament, is responsible for the generation of steady force and shortening, both in the sarcomere, the structural unit of the striated muscle cells, and in the cytoskeletal contractile units of all cells.
Here, we identify a myosin motor as the functional proteolytic fragment (heavy mero-myosin, HMM) of the muscle myosin II protein (either purified from an animal model or engineered), able to bind to actin, hydrolyse ATP, and generate force and movement. \\
The definition of the emergent properties of muscle myosin \textit{in vitro} has recently become accessible by means of a one-dimensional synthetic nanomachine based on the Dual Laser Optical Tweezers (DLOT) technology~\cite{pertici2018myosin}. 
The nanomachine, powered by myosin motors purified from the skeletal muscle and interacting with a single actin filament attached to a bead trapped in the focus of DLOT, is able to reproduce the collective mechanism of muscle myosin in the presence of physiological concentrations of ATP (2 mM)~\cite{pertici2018myosin,pertici2020myosin,pertici2021muscle}. 
In a recent work we designed a reverse engineering procedure to characterise the mechano-kinetic performance of a small ensemble of myosin motors, working collectively, in interaction with an actin filament~\cite{buonfiglio2024force}. 
In particular, we developed a stochastic framework, inspired by typical phenomenological model for the actin-myosin interactions~\cite{huxley1957muscle, woledge1985energetic}, to describe the output of the experiments conducted with the synthetic nanomachine under isometric conditions (i.e. without net relative sliding between actin and myosin array). 
Despite the bulk of data characterizing the mechano-kinetics of muscles at the cellular level, the definition of the corresponding microscopic parameters at the molecular level is still under investigation.
This is because inferring their definition from cells studies is challenging, due to the structural organization of the motors in the three-dimensional lattice of the sarcomere.
On the one hand, studies on the cellular level cannot resolve the details of this motor-coupling mechanism, which is difficult to isolate from the contribution of the other proteins. On the other hand, the collective dynamics as stemming from an ensemble of homologous molecules, cannot be predicted by single molecule experiments on purified proteins~\cite{finer1994single, ishijima1996multiple}.
Theoretical models adopted so far~\cite{buonfiglio2024force, buonfiglio2025resolving} are based on the assumption that the ensemble of motors is homogeneous. We refer to an ensemble as homogeneous when it shows a uniform population of motors, each one indistinguishable from the others in terms of mechanical parameters and kinetics. This assumption enables a successful investigation of the performance of a molecular motors array, and allows to quantify the performance of pure myosin isoforms. However, this assumption is not be accurate for several experimental scenarios of interest. 
In fact, the mechanical and kinetic performance of different types of striated muscles depends on the myosin II isoform expressed at the cellular level.
Specifically, the usual classification of the skeletal muscle cells (or fibres) in slow and fast-twitching fibres primarily depends on the myosin heavy chain (MyHC) isoform.
The MyHC is a subunit of the myosin protein responsible for the mayor enzymatic and mechanical properties, like the ATPase activity and the myosin-actin interaction.
Mammalian skeletal muscle contains four major MyHC isoforms~\cite{schiaffino2011fiber}: a slow isoform, MyHC-I, also knowns as MHC-$\beta$, and three fast isoforms, MyHC-, IIa, IIb, and IIx. These fibre types differ in both metabolic and force performance, and the predominance of one type or the other determines the function of the skeletal muscle. For example, slow fibres are abundant in postural muscles, also known as slow-twitching muscle, which have to contract for longer (exerting generally lesser force) and are resistant to fatigue. Fast-twitching muscle are usually responsible for voluntary movements, can exert a higher force, and are more prone to fatigue.
Although some mammalian skeletal muscles consist of a single fibre type, more commonly they contain a heterogeneous population of fibres, including intermediate hybrid fibres expressing specific combinations of two or more MyHC isoforms \cite{medler2019mixing}. 
Heterogeneous MyHC isoform composition is not restricted to skeletal muscle. 
In both atrial and ventricular muscle of small mammals the fast $\alpha$-MyHC is the dominant isoform. 
In large mammals the atrial myocardium, in addition to a majority of the faster myosin isoform, contains a certain proportion of the slow $\beta$-MyHC isoform~\cite{gorza1982myosin, SYROVY1989441, cummins1986myosin}, while the ventricle contains only the slow one. 
Isoform heterogeneity also occurs in non-muscle cells, where non muscle myosin isoforms (NMIIA, NMIIB and NMIIC) can co-assemble into heterotypic bipolar filaments containing motors with distinct mechano-kinetic properties, which may allow cells to tune force generation, tension maintenance, and cytoskeletal remodelling, according to local functional demands~\cite{BEACH20141160}.
Other interesting cases are the study of the effects of mutations in contractile proteins, mostly occurring in heterozygous form and therefore likely implying a condition of ensemble heterogeneity. 
In this regard, in a recent study~\cite{pertici2026mechanics}, the effects of the heterozygous p.E334Q mutation in cytoskeletal $\gamma$-actin have been characterised in relation to the mechanical performance with an HMM ensemble, to investigate the molecular alterations caused by the mutation that underlie the pathological phenotype. The p.E334Q mutation has been found to lead to alterations, generically classified non-muscle actinopathies, that mostly affect the cerebral cortex development and the muscle tone in humans~\cite{di2016update}.
In \cite{pertici2026mechanics} the dynamics of an HMM preparation interacting with a wild-type (WT) actin filament, which does not carry the mutation, were compared to the dynamics of the same HMMs preparation interacting with a mutant filament. 
Their performance was analysed within the framework of the stochastic model developed in \cite{buonfiglio2024force}, but such theoretical approach did not allow for the analysis of a preparation with a heterogeneous actin filament (heterofilament), in the specific case composed of $50\%$ mutant and $50\%$ wild-type monomers, reflecting the heterozygous condition of affected patients.
In this case, even though heterogeneity resides in the molecular track rather than in the myosin array, it determines distinct populations of acto-myosin complexes with distinct mechano-kinetic properties. 

Besides the experimental settings described above, a theoretical model able to describe the dynamics of a heterogeneous ensemble of myosin motors (or tracks), characterised by different mechanical and dynamical properties, is extremely relevant in the future application of synthetic nanomachines to HMM ensembles purified from mutant human samples or animal models, or from muscles that do not express a single myosin isoform.
Moreover, this approach is a valuable tool to determine whether a change in ensemble behaviour is primarily associated with the relative abundance of the two populations or with population-specific mechano-kinetic properties.
In this paper, we are going to concretely address this issue. 
A further improvement of the proposed method resides in a novel fitting procedure to provide an estimate of the system size $N$.
This is achieved with a parameter optimisation that relaxes the constraint of an \textit{a priori} fixed system size~\cite{buonfiglio2024force, buonfiglio2025resolving}, including this latter parameter among those inferred in the fitting process. The increased number of unknown parameters can be handled by exploiting the fact that the experimental conditions considered here are well-described by a simple phenomenological model of ATP-driven cyclic actin-myosin interactions. Such a simplified model involves a {\it single} force-generating state for the myosin motor instead of two states~\cite{buonfiglio2024force, buonfiglio2025resolving}. This assumption is appropriate when investigating the performance of a system designed to mimic sarcomere-level dynamics in mammalian skeletal muscle under room-temperature conditions. 

The other significant novelty of this work, and as anticipates above, is the fact that we addressed the complexity of a heterogeneous array of molecular motors involving the mechano-kinetic performance of two different populations of myosin molecules.
Specifically, we characterise the global force of a heterogeneous ensemble in terms of the probability distribution of two homogeneous arrays working independently one from the other. 
The working hypothesis of independent ensemble is justified by experimental evidence from most skeletal and cardiac muscles, in which multiple (slow and fast) myosin isoforms coexist within the tissue in different proportions, under the assumption that their interactions are not mutually antagonistic. 
The techniques proposed in this paper are expected to be employed to derive quantitative estimates of the mechano-kinetic parameters of individual actin-myosin interactions from the performance of heterogeneous ensembles of molecular motors, either purified from mutant or wild-type human samples and animal models or expressed recombinantly.

The paper is organised as follows: in Section \ref{sec: Model} the two-state phenomenological model for the actin-myosin interactions is presented in a stochastic framework. We consider a finite-size heterogeneous ensemble of fast and slow motors and we study the population dynamics of actin-bound motors in terms of microscopic attachment and detachment processes. Then, we characterise the global force exerted by the heterogeneous ensemble, obtaining an expression for the force probability distribution as a function of the model's mechano-kinetic parameters. The Results section (Section \ref{sec: Results}) is divided into four parts. In \ref{sec: Results hom ensemble} we describe an estimation procedure to characterise the mechano-kinetic parameters underlying the stochastic dynamics of an homogeneous ensemble of a finite, although unknown, size.
In \ref{sec: Results exp} we apply the estimation procedure for an homogeneous ensemble to experimental data with myosin motors purified from mammalian skeletal muscle. Then, we focus on the heterogeneous ensemble. In \ref{sec: Results het degree est} we detail the estimation procedure to determine the heterogeneity degree of the ensemble, in terms of the fraction of fast motors, in the case where the homogeneous populations of fast and slow motors can be separately investigated and characterised. Lastly, we consider a different scenario in which only the slow population ensemble can be directly investigated. Furthermore, in \ref{sec: Results het fast} we describe how to infer the fast population mechano-kinetic parameters from the analysis of an heterogeneous ensemble with known heterogeneity degree. An outlook of the possible applications with reference to the relevant physiological context is finally provided in the Discussion section \ref{sec:discussion}.

\section{The model} \label{sec: Model}

We consider a system composed of $N$ molecular motors that originate from two different populations, which we generically refer to as type $F$ motors (with kinetics corresponding to the fast myosin isoform) and type $S$ motors (with kinetics corresponding to the slow myosin isoform). More specifically, $N = N_F + N_S$, with $N_F = pN$ and $N_S = (1-p)N$, where $p$ is the fraction of fast motors. Each molecular motor interacts with the nearest actin monomer, following a kinetic scheme that describes the transitions between motor states, each one associated with a distinct spatial configuration and a different generated force. Following a simple phenomenological models for actin-myosin interactions~\cite{huxley1957muscle,woledge1985energetic}, we define a detached state $D$ for a motor that is not bound to the actin filament (therefore it exerts no force), and a force-generating attached state, A. Later on we will detail the force exerted by motors in this configuration. Each motor of the ensemble cyclically interacts with the actin filament according to the following kinetic scheme:
\begin{equation}
\label{eq: kinetic scheme}
\ce{D <=>[{k_+}][{k_-}] A}
\end{equation}
The transitions between the motors' states are governed by the kinetic rate constants $\vb{k}=(k_+, k_-)$, that represent the probability (per unit of time) for the single motor to jump from a state to another. 
This phenomenological model is adequate to describe dynamical regimes corresponding to high-temperature actin-myosin interactions in various mammalian striated muscle cells, such as rabbit fast and slow-twitch skeletal muscle~\cite{pertici2018myosin}. 
In such conditions the low-force, pre-working stroke A$_1$ configuration~\cite{huxley1971proposed} is a transient state, occupied by a negligible fraction of motors (on average). 
The choice of the two-state model is further discussed in the Appendix\ref{sec: App min mod validation} with the following approach: we simulated the dynamics of a phenomenological three-state model, like the one proposed in \cite{huxley1971proposed}, by assuming the kinetic rate constants fitted on experimental data sets on purified proteins from rabbit skeletal muscles (soleus and psoas). 
The fraction of motors in the A$_1$ configuration is quantified in typically lower than $10\%$, suggesting that the two-state model returns indeed a sound interpretative scenario, under these operating conditions. We also recall that the three-state model for actin-myosin interaction dynamics was studied in details in ~\cite{buonfiglio2024force, buonfiglio2025resolving}, with an adapted version of the methodologies here employed.

\subsection{Stochastic population dynamics of an heterogeneous ensemble}
\label{sec: masterEq}

\begin{figure}[htbp]
    \centering
    \begin{subfigure}[b]{0.49\textwidth}
        \centering
        \includegraphics[width=\textwidth]{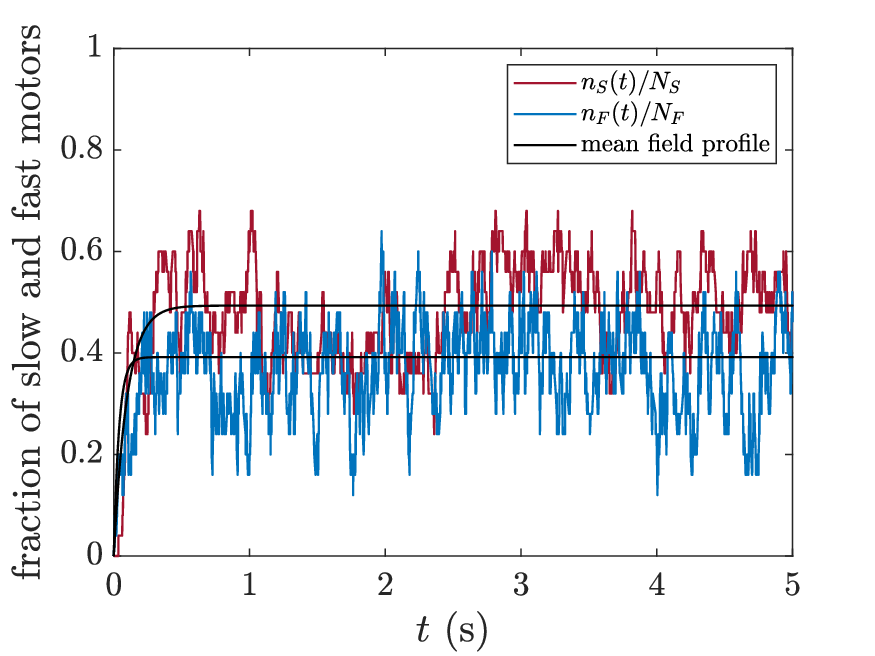}
        \caption{}
    \label{fig: population dynamics}
    \end{subfigure}
     \begin{subfigure}[b]{0.49\textwidth}
        \centering
        \includegraphics[width=\textwidth]{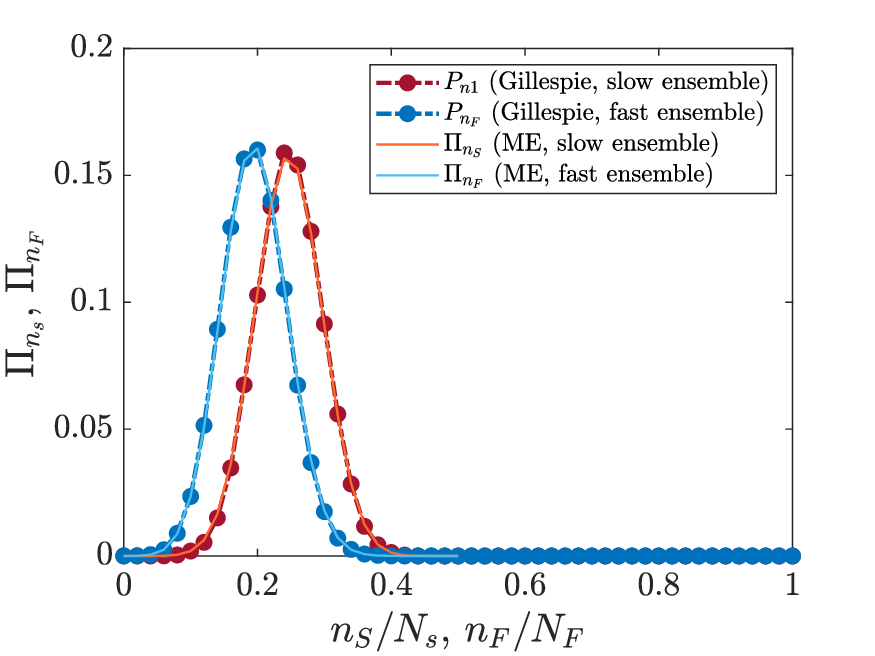}
        \caption{}
    \label{fig: stat state conc}
    \end{subfigure}
    \begin{subfigure}[b]{0.49\textwidth}
        \centering
        \includegraphics[width=\textwidth]{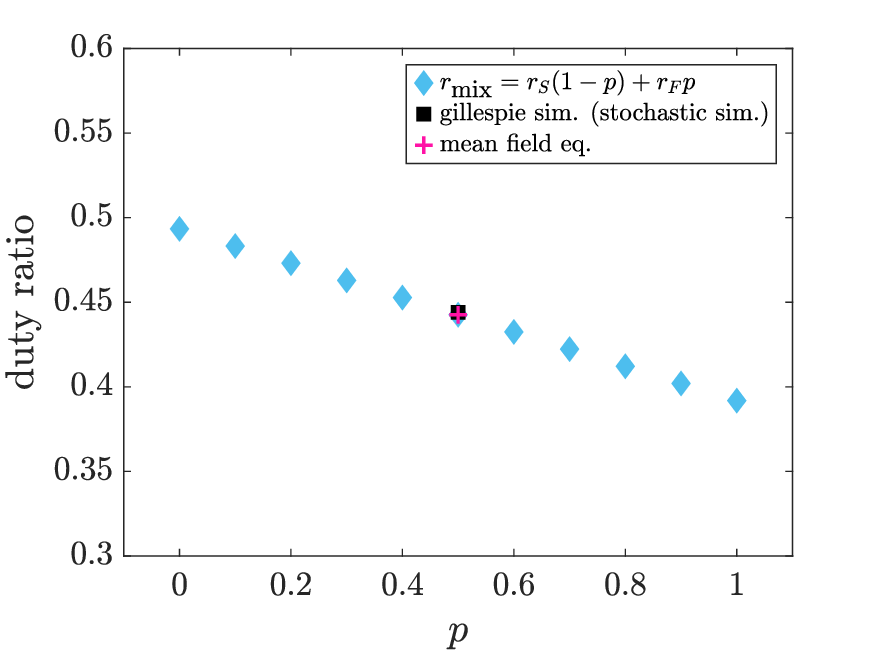}
        \caption{}
    \label{fig: duty ratio}
    \end{subfigure}
    \caption{\textbf{Population dynamics of the heterogeneous ensemble.} \\
    \textbf{Panel (a)}: time evolution of the fraction of motors in the fast $F$ and slow $S$ populations, for $N=50$ and $p=0.5$. The kinetic parameters that we considered are: $k_+^F= 10$, $k_-^F= 15.52$; $k_+^S= 5 $, $k_-^S= 5.13 $, all expressed in $\text{s}^{-1}$. Stochastic trajectories (solid coloured lines) are obtained with the Gillespie simulations, while black lines are obtained by integrating the mean field equations. 
    \textbf{Panel (b)}: comparison between the master equation (discrete) probability mass function $\Pi_{n_{\{F,S\}}}$ (solid lines, light colours) and its estimate $P_{n_{\{F,S\}}}$ (symbols, dark colours) obtained as the time-averaged 
    distribution at the stationary state of the stochastic trajectories.
    \textbf{Panel (c)}: duty ratio of the heterogeneous ensemble as a function of $p$. The light blue symbols are the theoretical prediction of Eq.~\eqref{eq: duty ratio}, with the same kinetic regimes and parameters adopted in the Panel (a). 
    The black square is the value of the duty ratio computed as the fraction of fast and slow attached motors, obtained from the stochastic simulations of the dynamics around the isometric plateau, while the red '+' marker is obtained from the mean field analysis.
    }
    \label{fig: popul dyn het}
\end{figure}

Let us initially focus on an homogeneous ensemble with only one population of motors,labelled with $j=\{F,S\}$. 
Following the single-motor kinetic scheme \eqref{eq: kinetic scheme} for the actin-myosin interaction, we consider the master equation for the probability distribution $P(\vb{n}_j(t),t)$ of the vector $\vb{n}_j(t)$ representing the system state at time $t$. 
The total number of motors is fixed; thus, the state of the system is univocally defined by specifying the number of motors in the force-generating configuration A, i.e., $n_j(t)$, while the number of detached motors is just $n_{0,j}=N_j-n_j$.
The master equation for the homogeneous ensemble reads:
\begin{equation}\label{eq: ME}
    \frac{\partial P(\vb{n}_j,t)}{\partial t} = \sum_{\vb{n}_j'\ne \vb{n}_j} \Bigl[ T(\vb{n}_j \vert \vb{n}'_j) P(\vb{n}'_j,t)- T(\vb{n}'_j \vert \vb{n}_j)P(\vb{n}_j,t) \Bigr],    
\end{equation}
where $T$'s denote the transition rates that are identified following the reaction cycle \eqref{eq: kinetic scheme}. Specifically, we have: 
\begin{itemize}
    \item 
    Attachment process: $\qquad T(n_j+1 \vert n_j)= k_+^j \ \frac{n_{0,S}}{N_j}= k_+^j \Big( 1- \frac{n_j}{N_j} \Big)$
    \item Detachment:  $\qquad T(n_j-1 \vert n_j)= k_-^j \ \frac{n_j}{N_j}$.
\end{itemize}
To determine the stochastic dynamics of the heterogeneous ensemble we consider the master equation of the heterogeneous system, of size $N$, with a state described by $\vb{n}(t)= (n_F(t), n_S(t))$ that incorporates the contribution of all the possible transition rates (attachment, detachment) for both populations. 
If we consider two non-interacting populations each transition rate acts only on one kind of motors, and the master equation of the heterogeneous system is the sum of the contribution of the transition rates for both populations:
\begin{equation}
\begin{split}
     \frac{\partial P(\vb{n},t)}{\partial t} & = \sum_{\vb{n}'\ne \vb{n}} \Bigl[ T(\vb{n} \vert \vb{n}') P(\vb{n}',t)- T(\vb{n}' \vert \vb{n})P(\vb{n},t) \Bigr] = \\
     & = T(\vb{n}|n_F-1) P(n_F-1,t) - T(n_F+1|\vb{n})P(\vb{n},t)+ \\
     & \quad + T(\vb{n}|n_F+1) P(n_F+1,t) - T(n_F-1|\vb{n})P(\vb{n},t) + \\
     & \quad + T(\vb{n}|n_S-1) P(n_S-1,t) - T(n_S+1|\vb{n})P(\vb{n},t)+ \\
     & \quad + T(\vb{n}|n_S+1) P(n_S+1,t) - T(n_S-1|\vb{n})P(\vb{n},t)
\end{split}
\end{equation}
where we have only highlighted the individual component that changes, according to the considered reaction 
Then we set forward to consider the Master equation in the more compact, matrix, form:
\begin{equation}
\label{eq: ME het}
     \frac{d P_{\vb{n}}(t)}{d t} = \sum_{n'} Q_{nn'} P_{n'}(t)    
\end{equation}
where the probability distribution is expressed as a time-dependent vector $P(t)$, with elements $P_{\vb{n}}(t)$, $\vb{n}=\{0, \dots, N\}$.
The matrix $Q$ with elements $Q_{nn'}=T(n|n')- \delta_{nn'} \sum_m T(m|n')$ denotes the infinitesimal generator that defines the instantaneous transition rates of the continuous-time Markov process, whose associated probability distribution evolves according to the master equation.
The stochastic dynamics of the attached motors can be simulated with the popular Gillespie algorithm~\cite{gillespie1976general,gillespie1977exact} (also known as the Stochastic Simulation Algorithm, SSA), which generates stochastic trajectories of the time evolution of the system state with probability distribution governed by the master equation.
In our simulations, the kinetic rate constants are chosen to mimic the isometric performance of the skeletal muscle myosin motors at room temperature, both for the fast and slow populations.
Figure~\ref{fig: population dynamics} shows the fluctuating population dynamics for an heterogeneous ensemble of $N=50$ motors, with a percentage of $50\%$ of fast motors (in blue), and, consequently, $50 \%$ of slow motors (in red). 
Selecting an initial condition with all the motors detached from the actin, after a transient the system populations evolve towards a stationary state, where the fraction of attached motors fluctuates around a stationary plateau. 
For the description of the average profile (the black dashed line) of the populations dynamics we refer to the mean field analysis in Section~\ref{sec: meanField}. 
Figure~\ref{fig: stat state conc} shows the comparison between the master equation (discrete) probability mass function $\Pi_{n_j}$, (solid lines, light colours) $ j \in \{F,S\}$ and its estimate $P_{n_j}$ (symbols, dark colours) obtained as the time-averaged distribution at the stationary state of the stochastic trajectories.
From the simulations of the stochastic dynamics around the stationary plateau we compute the average number of motors $n_F$ and $n_S$ in the force-generating state. Here, the duty ratio of the heterogeneous ensemble is defined as the average fraction of attached motors in the fast and in the slow populations, i.e., $r= (n_F+n_S)/N$. 
Indeed, the duty ratio of the single, homogeneous, population is $r_j=n_j/N_j$, with $j=\{F,S\}$, $N_F=pN$ and $N_S=(1-p)N$. Hence, for the heterogeneous ensemble, we can write
\begin{equation}
    \label{eq: duty ratio}
    r= \frac{r_S N_S + r_F N_F}{N} = (1-p) r_S + p r_F .
\end{equation}
This simple relation describes the value of the duty ratio for a mixed ensemble as a function of the fraction $p$ of fast motors in the mixture.
A visual reference of the duty ratio of the heterogeneous ensemble as a function of the fast motor fraction $p$ is shown in Figure~\ref{fig: duty ratio} (light blue symbols). 
The black marker in the figure is the values of the duty ratio of the heterogeneous ensemble computed from the synthetically generated dynamics using the Gillespie algorithm, for $p=0.5$.
The choice of the kinetic parameters corresponds to a dynamical regime in which the rate of force development mimics the ones of fast and slow myosin isoforms in a mammalian skeletal muscle.

\subsection{Force generated by the heterogeneous ensemble}
\label{sec: ensembleForce}

\begin{figure}[htbp]
    \centering
    \begin{subfigure}[b]{0.49\textwidth}
        \centering
        \includegraphics[width=\textwidth]{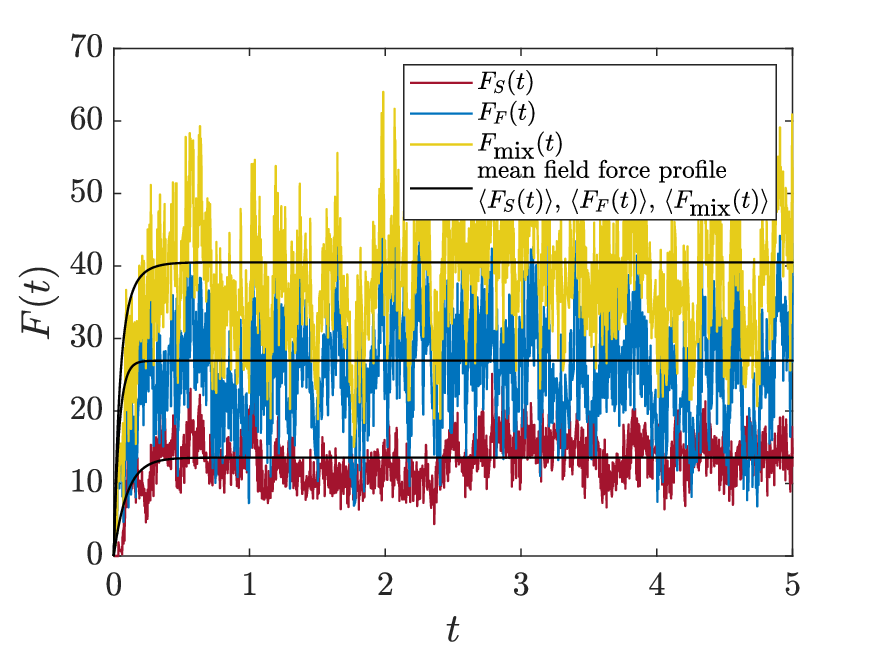}
        \caption{}
    \label{fig: ensemble force}
    \end{subfigure}
    \begin{subfigure}[b]{0.49\textwidth}
        \centering
        \includegraphics[width=\textwidth]{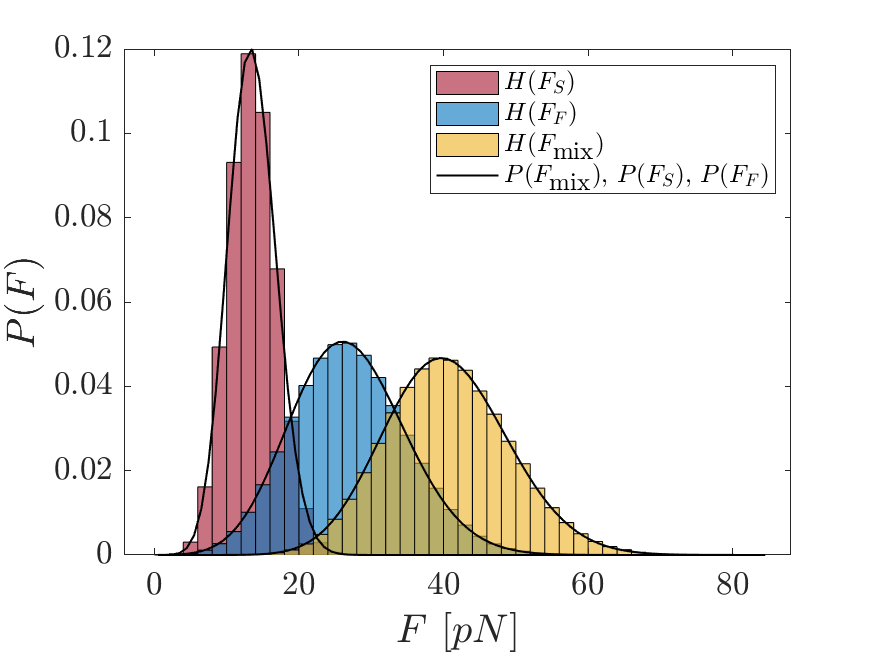}
        \caption{}
    \label{fig: pdf ensemble force}
    \end{subfigure}
    \caption{\textbf{Stochastic simulation of the heterogeneous ensemble force } \\
    \textbf{Panel (a)}: time evolution of the ensemble force for the fast $F$ (blue) and slow $S$ (red) populations, and for the mixed ensemble (yellow) of $N=50$ motors, with $p=0.5$. The single motor forces for the fast and slow populations are $f_0^F= 5 \text{ pN}$, $f_0^S= 2 \text{ pN}$, and the kinetic parameters adopted for the fast and slow populations are $k_+^F= 10$, $k_-^F= 15.52$; $k_+^S= 5 $, $k_-^S= 5.13 $, all expressed in $\text{s}^{-1}$. Stochastic trajectories (coloured) are obtained with the Gillespie simulations, while the solid black lines are the mean field force profile of the fast, slow, and heterogeneous ensembles.   
    \textbf{Panel (b)}: probability density functions $P(F_j)$ of the fast, slow, and heterogeneous ensemble force (black solid lines), superimposed on the empirical histograms $H(F_j)$ of the fluctuations obtained from the stochastic simulations. The mechano-kinetic parameters are the same adopted to generate the trajectory depicted in Panel (a). 
    }
    \label{fig: ensemble force het}
\end{figure}

Let us now compute the force generated by an ensemble of fixed size $N$, for a given value of the parameter $p$. We assign to each motor the force that it is able to exert: a detached motor does not exert any force, a fast (slow) motor in state A exerts a force $f_F$ ($f_S$). 
For each population $j$ (fast and slow), we define the force exerted by a motor in the state A during the working stroke to be $f_0^j$. 
As in \cite{buonfiglio2024force} we take into account the random orientation of the motors coating the support in the synthetic nanomachine, which progressively reduces the force developed during an actin-myosin interaction \cite{ishijima1996multiple}. Greater the deviation from the correct (i.e. \textit{in situ}) orientation, and larger the force reduction.
Specifically, for each motor, we assume that the force exerted by a motor in one ATPase cycle, $f$ is a random variable uniformly distributed over the interval $\mathcal{I}_j=[f_0^j/10,f_0^j ]$, therefore $f^j \sim \mathcal{U}(\mathcal{I}_j)$ (we refer the readers to \cite{buonfiglio2024force} and Fig.~7 in \cite{pertici2021muscle} for more detailed explanations of the technicalities underlying this assumption). Under the hypothesis that fast and slow motor populations operate independently as force generators, the total force exerted by the heterogeneous ensemble can be expressed as the sum of the force exerted by the two populations:
\begin{equation}
\label{eq: F mix}
        F(t)= F_F(t)+F_S(t)= F_{n_F} (t) + F_{n_S} (t)= \sum_{i=0}^{n_F(t)} f^F_i + \sum_{i=0}^{n_S(t)} f^S_{i}\,
\end{equation}
where the average number of motors $n^j(t)$, $j=\{F,S\}$ at any time $t$ are predicted by the stochastic simulations.   
Eq.~\eqref{eq: F mix} can be adopted to synthetically generate trajectories of the stochastic force of the heterogeneous ensemble.
An example of such a trajectory of the force is shown in Figure~\ref{fig: ensemble force} (yellow line), for an ensemble of $N=50$ motors, with a percentage of $50\%$ of fast motors, with a choice of the mechano-kinetic parameters (i.e., the single motor force and the kinetic rate constants) reported in the caption of the figure. The stochastic profiles in the figure reproduce the force development: starting from an initial condition in which all the motors are detached from the actin and the force is zero, due to the actin-myosin cyclic interactions of attachment-force generation-detachment, the total force attains a steady isometric plateau. 
The blue and red fluctuating trajectories refer to the force of the fast and slow ensembles respectively, 
as stemming from the dynamics portrayed in Figure~\ref{fig: population dynamics}. Notice that we have deliberately  imposed a lower value (with respect to the ensemble of fast motors) the typical single motor force for the ensemble of slow motors, with the aim of reproducing the functional differences between fast and slow myosin isoforms in skeletal muscle. The solid black lines identify the average force profile for the three ensembles (see the next subsection for details).

In Figure~\ref{fig: pdf ensemble force} the empirical distributions $H(F)$ of the force fluctuations around the isometric plateau are also shown. These distributions are computed as the histograms of the force trajectories, for the fast (blue bars), slow (red bars) and heterogeneous (yellow) ensembles. 
In Section~\ref{sec: forceDistro} we will detail how to obtain the theoretical probability distributions $P(F)$ of the forces (here depicted with solid black lines).

\subsection{Probability distribution of the ensemble force}
\label{sec: forceDistro}

The fluctuations of the ensemble force around the average isometric plateau are a manifestation of the mechano-kinetic parameters underlying the dynamics. Thus, we are going to characterise the probability distribution of the ensemble force at the isometric plateau in terms of the relevant parameters i.e. the single motor force and the attachment/detachment kinetic rate constants. 
In doing this, we generalise the expression of the probability distribution of the force fluctuations around the isometric plateau, for the homogeneous ensemble of $n$ force generating motors \cite{buonfiglio2025resolving}:
\begin{equation}
\label{eq: P(Fj)}
    P(F_{n}=\mathcal{F})= \sum_{q=0}^N P(F_{q}=\mathcal{F}) \Pi_{n=q},
\end{equation}
where $P(F_{q}=\mathcal{F})$ is the probability distribution of the force generated by $q$ motors in the force-generating state A, which is weighed with the probability of having $q$ motors in the force-generating state, i.e. the component of the stationary solution of the master equation \eqref{eq: ME} associated with the system the state $n$, $\Pi_{n=q}$. 

For an heterogeneous ensemble of size $N$ with a proportion $p$ of fast motors, the probability distribution of the force is
\begin{equation}
    P(F_{n_F,n_S}=\mathcal{F})= \sum_{q_F=0}^{pN} \sum_{q_S=0}^{(1-p)N} P(F_{q_F,q_S}=\mathcal{F}) \Pi_{(q_F,q_S)=(n_F,n_S)}
\end{equation}
In agreement with experimental results we can assume that (i) the sizes of the two populations of fast and slow motors are fixed, and (ii) fast and slow motors act independently from each others. 
Consequently, the solution of the master equation \eqref{eq: ME het} $P_{\vb{n}}(t) \equiv P_{n_F, n_S}(t)$ can be factorised in the two marginal probability distributions: $P_{n_F, n_S}(t)= P_{n_F}(t) P_{n_S}(t)$. 
Hence, around the isometric plateau, the probability distribution of the force fluctuations becomes
\begin{equation}
\label{eq: PF}
    P(F_{n_F,n_S}=\mathcal{F})= \sum_{q_F=0}^{N_F} \sum_{q_S=0}^{N_S} P(F_{q_F,q_S}=\mathcal{F}_F+\mathcal{F}_S) \Pi_{q_F=n_F} \Pi_{q_S=n_S}.
\end{equation}

The probability distributions of the force exerted by a fixed number of motors $F_q=\sum_{i=0}^q (f)^i$ is distributed according to a generalised Irwin-Hall distribution \cite{hall1927distribution} (the so called uniform sum distribution $\mathcal{S}$), of the form:
\begin{equation}
\label{eq: IH}
    \mathcal{S}_q (x)= \frac{1}{b-a} \ g(y;q) \qquad \qquad \qquad \text{with:} \qquad \qquad y=\frac{x-qa}{(b-a)}
\end{equation}
where:
\begin{equation*}
        g(y,q)= \frac{1}{2(q-1)!} \sum_{k=0}^n{(-1)^k \binom{q}{k} (y-k)^{q-1} \sgn(y-k)}
\end{equation*}
for the sum of $q$ random variables $x_i$ ($f_i$), each of them defined in the interval $[a,b]$.
The generalised Irwin-Hall distribution can be well-approximated with a Gaussian distribution if $q \gg 0$ (for more details, see Section 2 in \cite{buonfiglio2024force}). In particular, for a one-population ensemble, the Gaussian distribution for the force exerted by $q$ motors is $\mathcal{G}(\mu, \sigma^2)$ with $\mu= f_0(11/20)q$ and $\sigma^2= f_0^2 (27/400)q $~\cite{buonfiglio2025resolving}.
In the case of two populations of motors acting independently one from the other, the probability distribution of the total force is the convolution of the two probability distributions $P(F_{q_F}=\mathcal{F}_F)$ and $P(F_{q_S}=\mathcal{F}_S)$, i.e. $\big(P(F_{q_F}) \ast P(F_{q_S})\big)(F):=\int_{-\infty}^{\infty} P(F_{q_F}=\mathcal{F}_F) P(F_{q_S}=\mathcal{F}-\mathcal{F}_S) d \mathcal{F}$, which, when adopting the Gaussian approximation for the probability distribution of the force of a fast and slow ensemble acting independently one from the other, leads us to:
\begin{equation}
    P(F_{q_F,q_S}=\mathcal{F}_F+\mathcal{F}_S) \simeq \mathcal{G}(\mu_F+\mu_S,\sigma^2_F+\sigma^2_S) \equiv \mathcal{G}(\mu_{\text{mix}},\sigma^2_{\text{mix}})
\end{equation}
with mean and variance
\begin{equation}
\begin{cases}
    \mu_{\rm mix}(f_0^{S,F}, q_{S,F}) = \displaystyle{\frac{11}{20} (f_0^S q_S+ f_0^F q_F)} \\
    \\
    \sigma^2_{\rm mix} (f_0^{S,F}, q_{S,F}) = \displaystyle{(f_0^S)^2 \frac{27}{400} q_S + (f_0^F)^2 \frac{27}{400} q_F}.
\end{cases}
\end{equation}
As a result,
\begin{equation}
\label{eq: PF gauss}
    P(F_{n_F,n_S}=\mathcal{F})= \sum_{q_F=0}^{N_F} \sum_{q_S=0}^{N_S} \mathcal{G}(\mu_{\rm mix}, \sigma^2_{\rm mix}) \Pi_{q_F=n_F} \Pi_{q_S=n_S}.
\end{equation}
It is worth noting that $P(F_{n_F,n_S})$ is a function of the mechano-kinetic parameters of the model, including the force of the single fast and slow motors, the kinetic rate constants entering the solution of the master equation, as well as of the proportion $p$ of fast motors.
As anticipated above, in Figure~\ref{fig: pdf ensemble force} the solid black lines refer to the theoretical expression of $P(F)$, as computed above. This latter curves superimpose nicely to the histogram of the force distribution $H(F)$, traced after the stochastic simulations of the heterogeneous ensemble dynamics.

\subsection{The mean field analysis}
\label{sec: meanField}

In this section we study the dynamics of the heterogenous ensemble in the mean field limit, by solving a system of ordinary differential equations (ODE) for the fraction of fast and slow force-generating motors. 
We define the fraction of motors in configuration A, as determined from the governing master equation, as: $\vb{z}=(z_F, z_S)= \big(\frac{n_F}{N}, \frac{n_S}{N}\big)$, for the fast and slow ensemble respectively. 
The two-dimensional vector $\vb{z}$ univocally identifies the state of the ensemble for a choice of the parameter $p$, since the total number of motors $N$ is fixed.
The concentration of fast (slow) detached motors at any time is $x_0^{F(S)}(t)=N_{F(S)}-z^{F(S)}(t)$, with $N_F= pN$, $N_S=(1-p)N$.

The time evolution of the fraction of motors in each configuration is described on average by the following set of two ODE (for both fast and slow motors labelled with the index $j=\{F,S\}$):
\begin{equation}
\label{eq: MF}
    \displaystyle{\frac{dz_j}{dt}}= k_+^j (1-z_j)- k_-^jz_j= k_+^j- z_j(k_+^j+ k_-^j)
\end{equation}
In Eq.~\eqref{eq: MF}, the sets of kinetic constants for the fast and slow motors are $\vb{k_F}=(k_+^F, k_-^F)$ and $\vb{k_S}=(k_+^S, k_-^S)$, which follow the kinetic scheme of Eq.~\eqref{eq: kinetic scheme} for single-type motors.
Thus, in the mixed ensemble, the evolution of the fraction of motors in the force-generating configurations $\vb{z}$ can be written as
\begin{equation}
    \label{eq: MF mix}
    \frac{d\vb{z}}{dt}= J \vb{z} + \vb{b},
\end{equation}
with
\begin{equation}
\label{eq: J mix}
    J= \begin{pmatrix}
        -(k_+^F+k_-^F) & 0 \\
        0 & -(k_+^S+k_-^S)
    \end{pmatrix} 
    \qquad \text{and} \qquad \vb{b}=
    \begin{pmatrix}
    k_+^F \\
    k_+^S
    \end{pmatrix},
\end{equation}
The equilibrium solution $\vb{z}^*=(z_F^*, z_S^*)$ is obtained by imposing $d\vb{z}/dt=0$, and reads:
\begin{equation}
    z_j^*= \frac{k_+^j}{k_+^j+k_-^j} \qquad \textnormal{for } j=\{F,S\} .
\end{equation}
Eq.~\eqref{eq: MF mix} is a linear system that can be solved explicitly. 
We here assume an initial condition $z_F(0)=z_S(0)=0$ meaning that all the motors are detached from the actin. One gets:
\begin{equation}
\label{eq: MF sol}
    z_j(t)= \frac{k_+^j}{k_+^j+k_-^j}\big(1-e^{-(k_+^j+k_-^j)}\big)= z^*(1-e^{-t/\tau_j}) \qquad j=\{F,S\} .
\end{equation}

Integrating Eq.~\eqref{eq: MF mix} for a suitable choice of the mechano-kinetic parameters (see caption of Figure~\ref{fig: popul dyn het}), we obtain the solid dashed lines in Figure~\ref{fig: population dynamics}.
The duty ratio of the ensemble, in this deterministic framework, is computed with Eq.~\eqref{eq: duty ratio} from the fraction of motors in the force-generating states $z_F, z_S$. The black '$+$' markers in Figure~\ref{fig: duty ratio} represent the duty ratio of the heterogeneous ensemble.
In this mean field context we define the flux through the attachment-detachment cycle as $\phi_j= z_j^* k_-^j= (k_+^j k_-^j)/(k_+^j + k_-^j)$ which is a proxy of the ATP consumed by the system.

We reproduce the behaviour of the force ensemble by considering the average number of motors in the force-generating configuration A for both the fast and slow populations. 
The average force of each population is computed by assigning the average force generated in the state A: $f_j$ for $j=\{F,S\}$. 
From the heterogeneous ensemble force, defined in \eqref{eq: F mix}, we obtain the average isometric force $\braket{F(t)}$ so that
\begin{equation}
\label{eq: F mean field}
        \braket{F(t)} = \braket{F^S(t)} + \braket{F^F(t)} = \braket{n_F(t)} \overline{f_F} + \braket{n_S(t)} \overline{f_S},
\end{equation}
where the average values of $f_j$ in the intervals $\mathcal{I}_j$ are denoted with the overline $\overline{\ \cdot \ }$. 
Since $\braket{n_j(t)}=N_j \braket{z_j(t)}$, we use \eqref{eq: MF sol} and obtain:
\begin{equation}
    \label{eq: Fj MF}
    \braket{F^j(t)}= N_j \braket{z_j(t)} \overline{f_j}= N \overline{f_j} \frac{k_+^j}{k_+^j+k_-^j} (1-e^{-(k_+^j+k_-^j)t})    
\end{equation}
for $j=\{F,S\}$.
In the stationary state we define the equilibrium fraction of motors as $\lim_{t \to \infty}\braket{z_j(t)} = z_j^*$, and $\braket{F(t)}$ converges at the isometric plateau $F_0$ of the force (for each homogeneous ensemble):
\begin{equation}
\label{eq: F0j}
    \begin{split}
        F_0^j &= N_j \overline{f_j} \frac{k_+^j}{k_+^j+k_-^j}  
    \end{split}
\end{equation}
where we exploit the fact that $\overline{f_j} = (11/20)f_0^j$ ($j=\{F,S\}$).
For the heterogeneous ensemble we obtain:
\begin{equation}
\label{eq: F0}
    \begin{split}
        F_0 &= (n_F)^* \overline{f_F} + (n_S)^* \overline{f_S} = r_F N_F \overline{f_F} + r_S N_S \overline{f_S}= \frac{11}{20} N \Big[ p r_F f_0^F + (1-p) r_S f_0^S \Big] \ .
\end{split}
\end{equation}

In Figure~\ref{fig: ensemble force} the average profiles of the force are shown as dashed lines, for the heterogeneous ensemble $\braket{F(t)}$, as well as the fast $\braket{F_F(t)}$ and slow $\braket{F_S(t)}$ populations. Starting from an initial condition in which all the motors are detached from the actin and the force is zero, the total force reaches a steady isometric plateau, whose value is given by Eq.~\eqref{eq: F0}. 
The choice of the kinetic parameters for the fast and slow populations corresponds to a dynamical regime that mimics the performance of skeletal myosin isoforms at room temperature. Therefore, the force development of the homogeneous ensemble can be reproduced with a single exponential function with a rate of force development $1/\tau_j= k_+^j+ k_-^j$ and asymptotic plateau $F_0^j$ for $j=\{F,S\}$. 
Using the latter, the average profile \eqref{eq: F mean field} of the force development for the heterogeneous ensemble is
\begin{equation}
    \braket{F(t)} = N \Big[ p \frac{k_+^F}{k_+^F+ k_-^F} \overline{f_F} \big(1-e^{-t/\tau_F}\big) + (1-p) \frac{k_+^S}{k_+^S+ k_-^S} \overline{f_S} \big(1-e^{-t/\tau_S}\big) \Big] \ .
\end{equation}
From Eq.~\eqref{eq: F mean field}, given a force profile $\braket{F(t)}$, we observe that $N$ and $p$ are proportional. In particular, the product $pN$ lies on a hyperbolae. In the stationary state, at the isometric plateau where $\braket{F(t)}$ approach the asymptotic value $F_0$, we get:
\begin{equation}
    \label{eq: MF hyp}
    N= \frac{F_0 (20/11)}{p r_F f_0^F+ (1-p) r_S f_0^S} = \frac{A}{p}-B
\end{equation}

\section{Results} 
\label{sec: Results}

\subsection{Mechano-kinetic parameter estimation: homogeneous ensemble}
\label{sec: Results hom ensemble}

We here detail the procedure adopted to characterise the mechano-kinetic performance of a small homogeneous ensemble of motors. 
Our aim is to provide an estimate of the relevant parameters underlying the actin-myosin interaction dynamics from a fitting procedure of the stochastic time series that exploits the model information as introduced above.
We validate the estimation procedure with both synthetically generated data, and experimental data taken from Ref.~\cite{buonfiglio2024force} (see Section \ref{sec: Results exp}).
The procedure (for a single data set and for the setting where data to be fitted are generated by numerical simulations) is as follows:

\begin{enumerate}
    \item 
    we generate the time series of the isometric force development for the homogeneous ensemble, via the Gillespie algorithm, implementing the dynamics described in Section~\ref{sec: masterEq} and~\ref{sec: ensembleForce}. 
    As initial condition for the stochastic population dynamics, we assume that all the motors are detached from the actin filament, i.e., $n(t=0)=0$. 
    The mechano-kinetic parameters are chosen to mimic the characteristics of the actin-myosin interactions of a slow skeletal muscle at room temperature (see Figure~\ref{fig: popul dyn het} and Figure~\ref{fig: ensemble force het}).
    The stochastic profile of the ensemble force in the figures (synthetically generated) is used to test the estimation procedure.
    
    \item 
    We fix the rate $1/\tau$ of force development of the ensemble, as the characteristic time of the exponential growth of the ensemble force's time series. This rate is fitted with the Eq.~\eqref{eq: Fj MF}. From the mean field analysis: $1/\tau_j= k_+^j+k_-^j$. 
    
    \item 
    We then remove a fixed transient time from the force trajectory, leaving only the portion of the trajectory at the stationary state, and we compute the empirical distribution of the ensemble force fluctuating around the isometric plateau $H(F)$. 
    
    \item We perform an optimisation procedure based on a simulated annealing algorithm \cite{cardoso1996simplex}. 
    The procedure is essentially a minimisation of a loss function computed as the distance between the empirical probability distribution $H(F)$ and the analytical counterpart $P(F)$. 
    The latter is expressed as a function of the model parameters $(N, f_0, \vb{k})$, calculated with relation \eqref{eq: P(Fj)}. 
    Specifically, deriving the $P(F)$ requires to solve the master equation \eqref{eq: ME het} in the stationary state, which provides the stationary probability distributions $\Pi_{n=q}$ of the generic system state $n$. 
    These terms contribute to the calculation of $P(F)$ as weighting factors for the probability distributions of the force exerted by a fixed number of motors, i.e. $P(F_{q_F,q_S} = \mathcal{F})$.
    The resulting probability distribution $P(F)=P\big(F(N, f_0, \vb{k})\big)$ of the ensemble force \eqref{eq: P(Fj)}, which depends on the model parameters, is compared to the empirical histogram $H(F)$ of the force trajectory fluctuating around the isometric plateau, obtained from the data set.
    The procedure converges when a minimum of $\mathcal{L}$ is found, for a selected interval of parameters values, and the corresponding parameters are the best fit parameters selected by the procedure.
    
    \item Each iteration of the estimation via the inverse scheme selects a set of the best fit parameters $(\hat{N}, \hat{f_0}, \hat{\vb{k}})$. 
    To improve the robustness of the estimates, the procedure is repeated multiple times and the estimated parameters were averaged over independent realisations.    
    We refer the readers to Appendix \ref{sec: App optimisation proc} for more details on the optimisation algorithm for an homogeneous ensemble of interacting motors.
\end{enumerate}

\begin{figure}[htbp]
    \centering
    \begin{subfigure}[b]{0.49\textwidth}
        \centering
        \includegraphics[width=\textwidth]{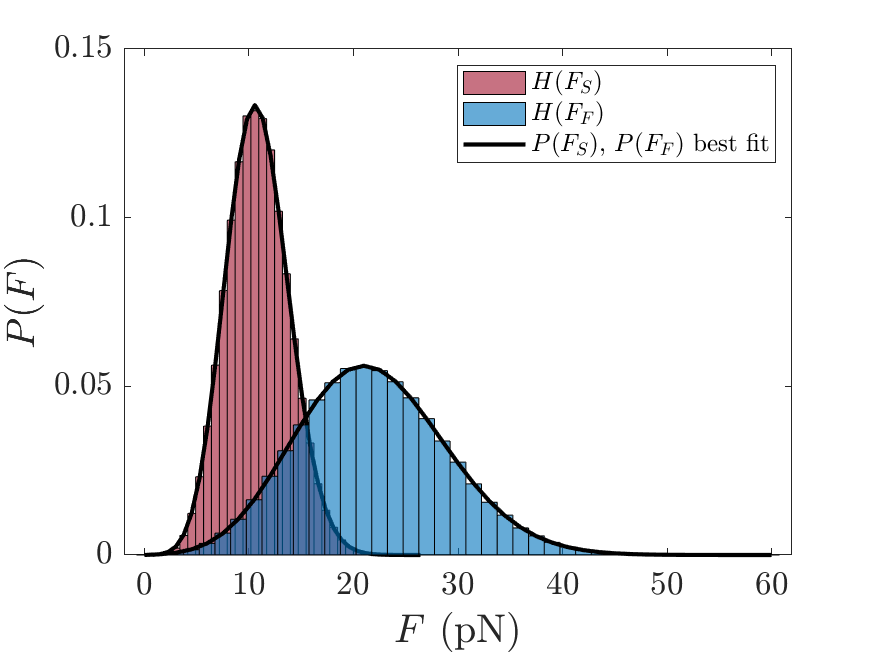}
        \caption{}
    \label{fig: fitData slow and fast hom}
    \end{subfigure}
      \begin{subfigure}[b]{0.49\textwidth}
        \centering
        \includegraphics[width=\textwidth]{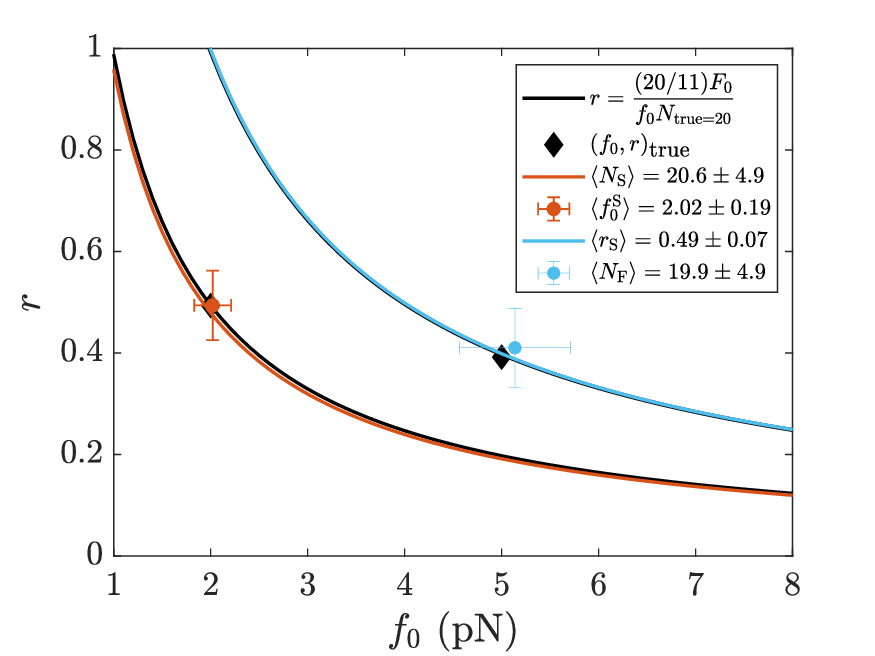}
        \caption{}
    \label{fig: f0VSr SlowFast}
    \end{subfigure}
      \begin{subfigure}[b]{0.49\textwidth}
         \includegraphics[width=\textwidth]{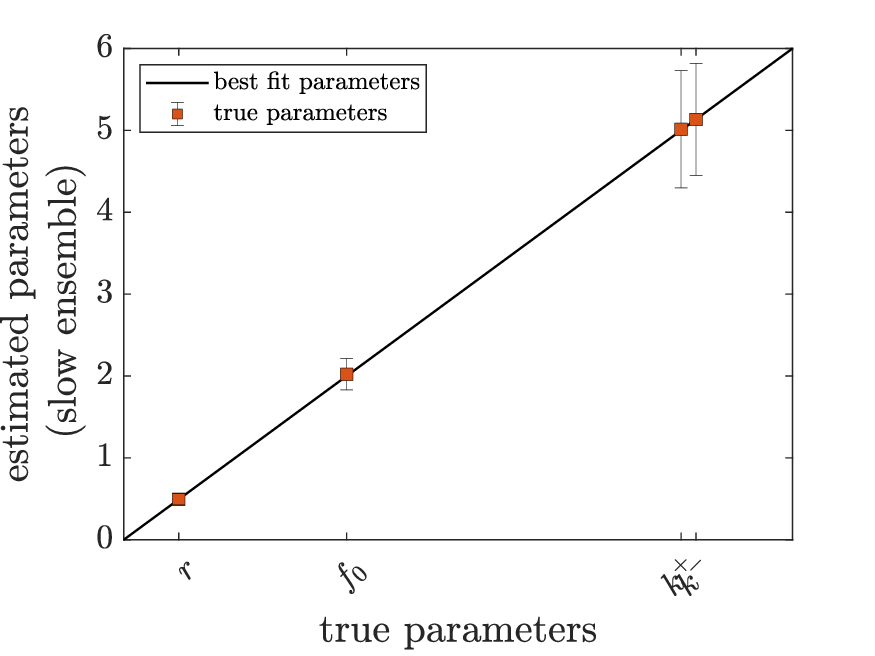}
        \caption{}
    \label{fig: kinParBest slow}    
    \end{subfigure}
     \begin{subfigure}[b]{0.49\textwidth}
        \centering
        \includegraphics[width=\textwidth]{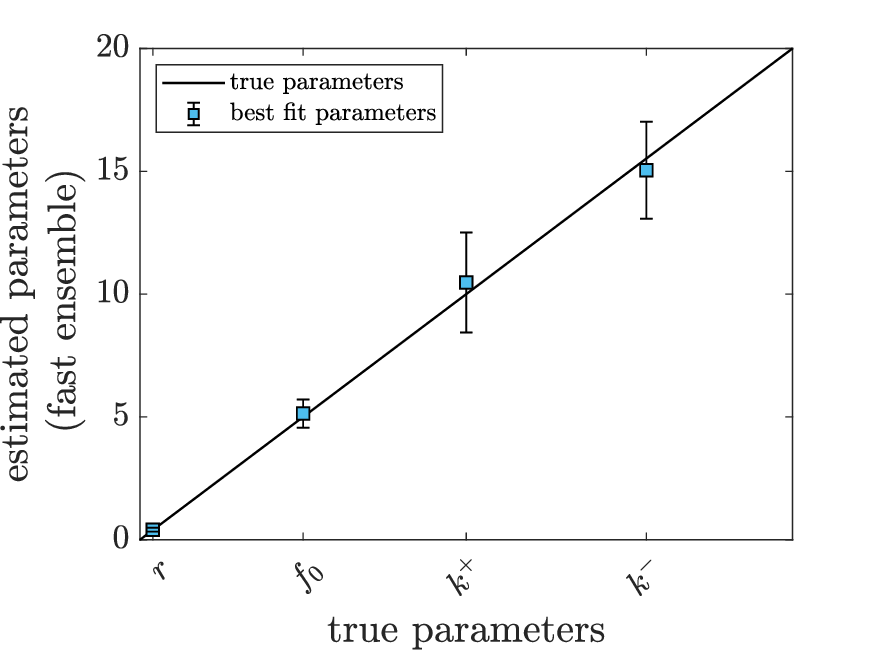}
        \caption{}
    \label{fig: kinParBest fast}
    \end{subfigure}
    \caption{
    \textbf{Estimation procedure results on individual homogeneous ensemble (fast and slow)} \\
    \textbf{Panel (a)}: the histograms $H(F_j)$, $j=\{\mathrm{F,S}\}$ are independently generated from synthetic data of the force around the isometric plateau, for an heterogeneous ensemble of $N=20$ molecular motors (fast population in blue, slow population in red), with the same kinetics of the one in figure~\ref{fig: ensemble force}. The black lines are the best fit $P(F_j)$ obtained with the optimisation procedure separately applied on the homogeneous ensembles.
    \textbf{Panel (b)}: in the plane $(f_0,r)$ are plotted the results of the estimation procedure for the fast (in light blue) and the slow (in light red) ensemble, the symbol and the error bars are the mean and SD computed from multiple independent realizations of the scheme.
    \textbf{Panels (c)},\textbf{(d)}: the mechano-kinetic parameters as selected by the estimation are plotted against their true values i.e., the ones set to generate the data, for the slow and fast ensemble respectively.
    }
    \label{fig: fit homog ensemble}
\end{figure}

In step 4 it is described the inverse scheme that aims to estimate the unknown model's parameters, by analysing the observable ensemble force.
The results of the procedure, applied both on a fast and on a slow homogeneous ensemble (separately), are displayed in Figure~\ref{fig: fit homog ensemble}.
In Figure~\ref{fig: fitData slow and fast hom} the empirical histograms $H(F_F)$, $H(F_S)$ are shown, with the best fit $P(F_F)$ and $P(F_S)$ superimposed in black, as the result of the optimisation procedure independently performed on the two homogeneous ensembles. 
In Figure~\ref{fig: f0VSr SlowFast} the results of the optimisation scheme is shown in the plane $(f_0,r)$ for both the fast (in light blue) and slow (in light red) ensemble. 
The light red and light blue symbols are the mean and standard deviation (SD) of multiple independent realisations of the estimation procedure, for the slow and fast ensemble, respectively. 
Then the hyperbolae are plotted employing the formulas of the mean field analysis for the homogeneous ensembles $r_j= F_0^j/(N_jf_j)$, $j=\{F,S\}$, with the best fit value of $N_j$ selected by the estimation procedure.
The best fit mechano-kinetic parameters are plotted in Figure~\ref{fig: kinParBest slow} and Figure~\ref{fig: kinParBest fast}, against their true value, i.e., the ones set to generate the data.
The symbols and error bars are the same as in Panel (b).
The values of the estimated parameters, as well as the system size $N$ selected by the procedure, are displayed in Table~\ref{tab: fit homogen} for the two homogeneous ensembles.
\begin{table}[h!]
	\begin{center}
		$\begin{array}{cccccc}
			\toprule
			\text{} & N & f_0 \ (\textnormal{pN}) &  r & \phi \ (\textnormal{s}^{-1}) & F_0 \ (\textnormal{pN}) 
			\\
			\midrule
			\text{True parameters}& 20 & 5 & 0.39 & 6.1 & 21.8 \\
			\text{FAST ensemble}& & & & &\\
			& & & & & \\
			\text{Estimated par. }& 19.9 \pm 4.9 & 5.1 \pm 0.6 & 0.41 \pm 0.08 & 6.0 \pm 0.3 & 21.8 
			\\
			\text{FAST ensemble }& & & & &\\        
			\midrule
			\text{True parameters}& 20 & 2 & 0.49 & 2.5 & 10.9
			\\
			\text{SLOW ensemble}& & & & &\\
			& & & & & \\
			\text{Estimated par. }& 20.6 \pm 4.9 & 2.02 \pm 0.19 & 0.49 \pm 0.07 & 2.49 \pm 0.08 & 10.9 
			\\
			\text{SLOW ensemble }& & & & &\\
			\bottomrule
		\end{array} $
    \caption{\textbf{Estimated parameters via the inverse scheme fed with simulated data, for each homogeneous ensemble of fast and slow motors.}\\
	The parameters are: the system size $N$, the force of a single motor $f_0$, the duty ratio of the ensemble $r$ and the rate of transition through the attachment–detachment cycle $\phi$. Mean and SDs are computed from different independent realisations of the optimisation procedure. Errors are below $10^{-3}$ if not explicitly provided.
		}
		\label{tab: fit homogen}
	\end{center}
\end{table}

\begin{figure}[htbp]
    \centering
    \begin{subfigure}[b]{0.49\textwidth}
        \centering
        \includegraphics[width=\textwidth]{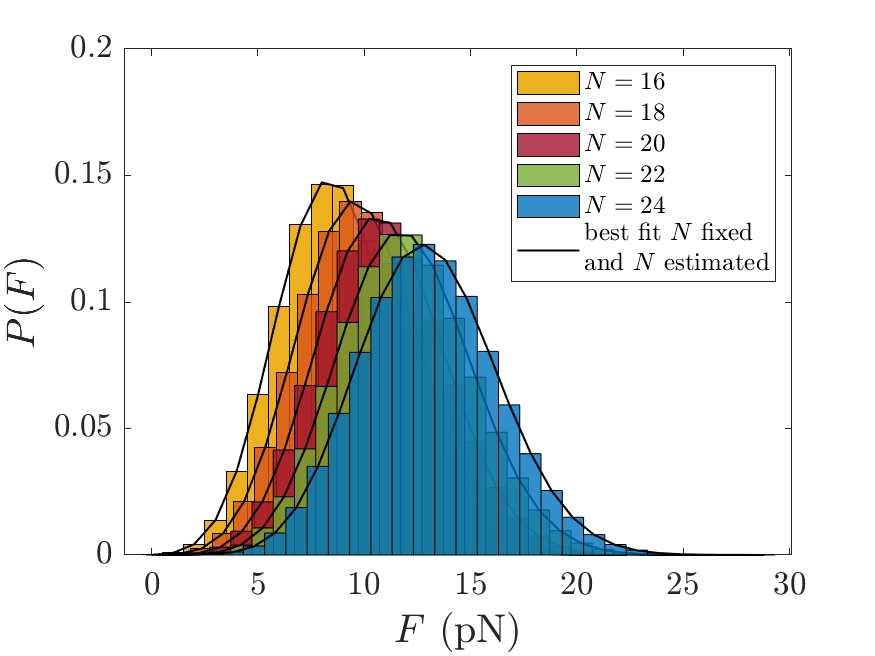}
        \caption{}
    \label{fig: fitData N est}
    \end{subfigure}
      \begin{subfigure}[b]{0.49\textwidth}
        \centering
        \includegraphics[width=\textwidth]{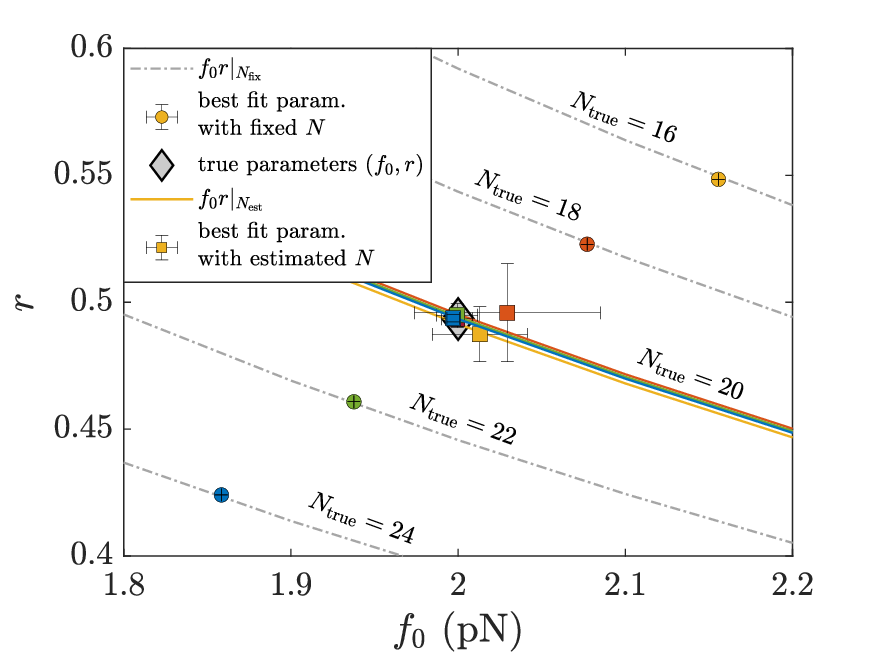}
        \caption{}
    \label{fig: f0VSr all}
    \end{subfigure}
    \caption{\textbf{Test of the performance of the estimation of $N$}\\
    \textbf{Panel (a)}: the coloured histograms $H(F^i)$ are generated from synthetic data of the force around the isometric plateau, for five homogeneous ensemble of different size $N_i=\{16$ (yellow), $18$ (orange) , $20$ (red), $22$ (green), $24$ (blue)$\}$, with the same kinetics and motor force of the slow ensemble in Figure~\ref{fig: ensemble force}. 
    The black lines are the best fit $P(F^i)$ obtained with the optimisation procedure separately applied on each homogeneous ensembles, with $N_i$ as a free parameter.
    \textbf{Panel (b)}: in the plane $(f_0,r)$ are plotted the results of both the estimation schemes with fixed and estimated $N$. 
    The big gray diamond represents the position of the true parameters in the plane $(f_0,r)_{\mathrm{true}}$. 
    Dashed gray lines are the hyperbolae on which falls the product $f_0 r$, with fixed system size $N=20$: each data set has a different distance from the origin of the plane. 
    The solid circles are the best fit parameters $(f_0,r)_i$ obtained with the procedure that fixes $N=20$ for each sets.
    Coloured solid lines represent the hyperbolae $f_0 r=f(F_0^i, N_i)$ with estimated $N$ provided by the estimation scheme.
    Coloured squared symbols are the best fit parameters selected by the procedure that consistently estimates $N_i$ for each data set. 
    }
    \label{fig: N fix vs N est}
\end{figure}

We aim in particular to provide an estimation of the system size $N$, which was considered as a fixed parameter in previous works from the authors~\cite{buonfiglio2024force, buonfiglio2025resolving}. 
In the cited works we described an experimental protocol that is able to estimate the total number motors available for the interaction with the actin filament. Essentially, the number of myosin molecules on the setup support surface that are able to interact with an actin monomer of the filament is determined by counting the number of mechanical rupture events when the myosin ensemble, brought into contact with the actin filament in an ATP-free solution to form rigor bonds is first displaced orthogonally to the actin–myosin interface, and then pulled away along the filament axis at a constant velocity, causing the rigor bonds to detach one by one; see also Results in \cite{buonfiglio2024force} for a more detailed description of the experimental protocol. 
Since the myosin molecule is a dimer, with two functional head domains, and under the assumption that in rigor both heads are attached to the filament \cite{reconditi2003conformation} number of available motors in $2$ mM ATP is twice the number $l$ of rupture events in rigor, $N= 2l$.
The value obtained as the average number of ruptures events per interaction in rigor conditions is then assumed as the nominal value in the theoretical modelling of the study. 
Since in principle each experimental data set could have a slightly different system size, which reflects on a slightly different isometric force of the ensemble (for example the rupture events distribution follows a Gaussian distribution with average $l= 7.9 \pm 1.1$ for the HMMs purified from
soleus muscle and disposed on the support at saturating concentrations  \cite{buonfiglio2024force}), fixing $N$ to a single value for all data set could results in different parameters estimation. \\
In the present work we refined the theoretical analysis by implementing a scheme that is also able to estimate, among the other target fitting parameters, the number of available motors, i.e., the system size.
In this way, we are able to provide an estimate of the system size that will be compared with that made experimentally accessible by the protocol discussed above.
Specifically, we constrain the value of $N$ to be compatible with the mean field analysis of the ensemble force, following from relation \eqref{eq: Fj MF}:
\begin{equation}
    \label{eq: N hom}
     N= \frac{F_0}{\overline{f}} \frac{k_+ + k_-}{k_+}
\end{equation}
where $\hat{\overline{f}}$, $\hat{k_+}$, $\hat{k_-}$ are the best fit parameters selected by the algorithm, and $F_0$ is the mean field ensemble force.
We then impose that $N \in \mathbb{N}$: as it represents the actual number of motors, the estimated quantity should be approximated to its closest integer value. 
This parameter is to be estimated together with the mechano-kinetic ones.
To test the performance of the fitting scheme, specifically with reference to the computed estimate for $N$, we generated a sufficiently large set of  synthetic data  for the stochastic ensemble force, each one with a different number of motors, i.e. $N=(16,18,20,22,24)$ and then applied the estimation scheme either with (i) a fixed system size $N_{\textnormal{fix}}$ or with (ii) $N$ as a free parameter. 
The results of the comparison is shown in Figure~\ref{fig: N fix vs N est}. The higher the number of free parameters in the procedure the larger the dispersion of the obtained solutions. However,  the estimated mean for $(f_0, r)$ fall close to the true nominal values (i.e. to the ones set in the generation of the synthetic data).

In Figure~\ref{fig: fitData N est} the coloured histograms $H(F_i)$ are generated from synthetic data of the force around the isometric plateau, for five homogeneous ensemble of different size $N_{\mathrm{true}}^i=\{16$ (yellow), $18$ (orange), $20$ (red), $22$ (green), $24$ (blue)$\}$, with the same kinetics and motor force of the slow ensemble in Figure~\ref{fig: ensemble force}. 
The black lines represent the best fit $P(F_i)$ obtained with the optimisation procedure separately applied on each homogeneous ensemble, with $N_i$ as a free parameter. These curves overlap with the best fit $P(F_i)$ obtained with fixed system size $N_i=20 \quad  \forall \ i=\{1, \dots ,5\}$, meaning that both the procedures correctly converges to good fit of the probability distribution.
In Figure~\ref{fig: f0VSr all} the results of the two different estimation schemes (with fixed and estimated $N$) are plotted in the parameter plane $(f_0,r)$. 
The large gray diamond represents the position of the true parameters in the plane $(f_0,r)$. 
Dashed gray lines are hyperbolae described by the mean field relation $f_0 r= (11/20)F_0^i/N_{\mathrm{fixed}}$ for the $i$-th data set (with size $N_{\mathrm{true}}^i$ defined above), computed with fixed $N=20$. 
Each curve has a different distance from the origin of the plane, which depends on the isometric force $F_0^i/N_{\mathrm{fix=20}}$ of the $i$-th data set, obtained as the average ensemble force of the $i$-th trajectory at the stationary state.
The solid circles are the best fit parameters $(f_0,r)_i$ obtained with the procedure that sets $N=20$ for all data sets.
Coloured solid lines represent the relation $f_0 r= \frac{(11/20)F_0^i}{N_{\mathrm{est}}}$ with estimated $N$ ($N_\mathrm{est}$) provided by the fitting scheme: the data sets cluster in a small region due to the fact that their distance from the origin of the plane depends on $F_0^i/N_i$,  $N_i$ being the estimated quantity.
Coloured squared symbols are the best fit parameters selected by the procedure that consistently estimates $N_i$ for each data set. 
Despite the larger error bars of these results they are all consistently closer to the true value of the parameters.

\subsection{Application of the estimation procedure to experimental data}
\label{sec: Results exp}

We here apply the estimation scheme described in Section \ref{sec: Results hom ensemble} to experimental data taken from \cite{buonfiglio2024force}. 
The data were acquired with the synthetic myosin-based nanomachine that mimics actin-myosin interactions and is able to reproduce isometric contractions under a specific length-clamp control \cite{buonfiglio2024force}. 
The machine was powered with HMM fragments purified from rabbit skeletal muscle. 
In the mentioned work we compared the performance of two different myosin isoforms, the (slow) type I (purified from rabbit soleus) and the (fast) type II (from rabbit psoas), characterising the mechano-kinetic parameters that underlie the isometric force development and the isometric plateau. 
The analysis was conducted with a fitting scheme that assumes a fixed system size, $N_{\mathrm{fix}}=16$, corresponding to the average number of motors available for the interaction with the actin filament. The average number of motors is experimentally estimated by counting the number of bond ruptures in ATP-free solution.

\begin{figure}[htbp]
    \centering
    \begin{subfigure}[b]{0.49\textwidth}
        \centering
        \includegraphics[width=\textwidth]{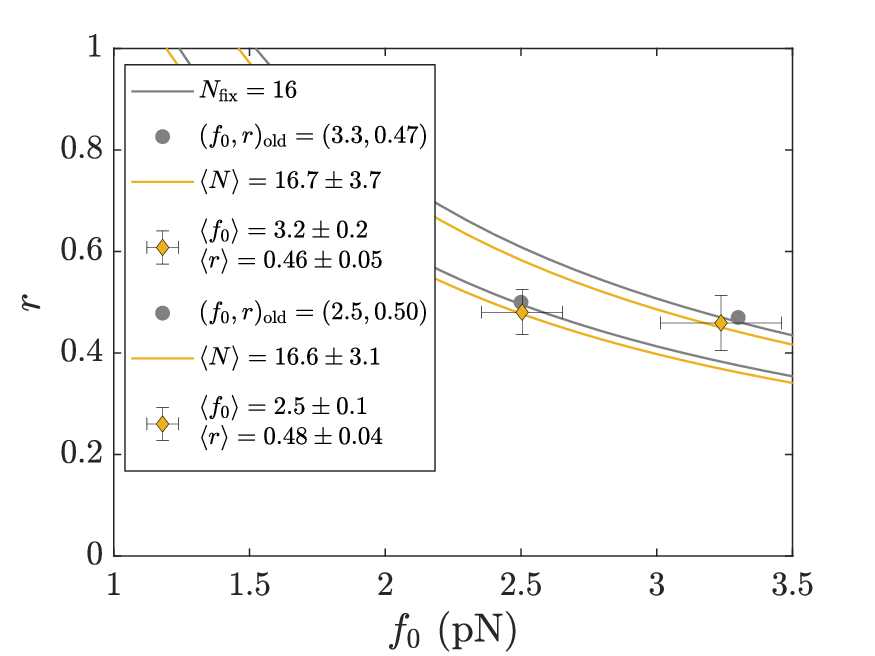}
        \caption{}
    \label{fig: par fit Soleus}
    \end{subfigure}
    \begin{subfigure}[b]{0.49\textwidth}
        \centering
        \includegraphics[width=\textwidth]{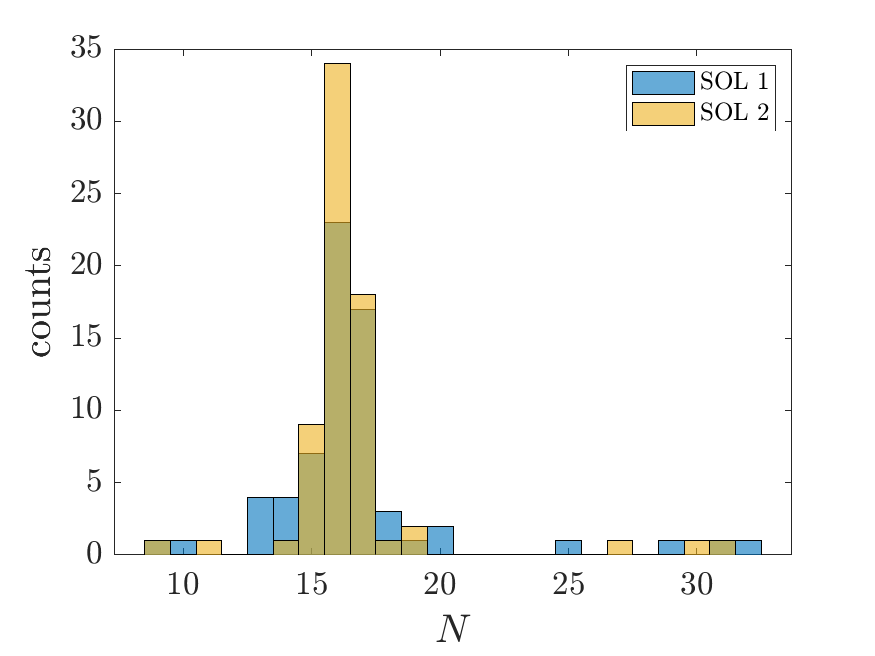}
        \caption{}
    \label{fig: N fit Soleus}
    \end{subfigure}
    \caption{\textbf{Validation of the estimation scheme with experimental data.}\\
    The estimation scheme for the homogeneous ensemble with variable system size is applied on two experimental data sets. The data are obtained with a synthetic machine powered by HMMs purified from rabbit soleus muscle. 
    \textbf{Panel (a)}: in the plane $(f_0,r)$, we compare previous results obtained with a fixed size, with the new estimates. 
    The mean field hyperbolae with fixed $N$ are the gray solid lines, while the gray dots represent the previous (mean) estimates derived in \cite{buonfiglio2024force}. The yellow lines and symbols, with error bars, are the new estimates obtained without fixing the size $N$. 
    \textbf{Panel (b)}: the size $N$ was computed from the mean field relation, rejecting values outside a physiological interval $[8,32]$ for myosin motors.
    }
    \label{fig: fit homog ensemble Soleus}
\end{figure}

Here we do not assume a fixed value for $N$ for the experimental data sets, accounting for the fact that different experimental records could present a slightly different number of available motors.
Following the optimisation procedure detailed in the Appendix~\ref{sec: App optimisation proc}, for any data set, we compute $N$ with equation \eqref{eq: N hom}, and we select the closest integer value.
In Figure~\ref{fig: fit homog ensemble Soleus} we compare the results in \cite{buonfiglio2024force} for the mechano-kinetic parameters of the slow ensemble (myosin motors purified from rabbit soleus) with $N=16$, with the new estimates obtained employing the estimation scheme for the homogeneous ensemble with unknown size $N$.
Two data sets are analysed: SOL1 in blue and SOL2 in yellow.
In Figure~\ref{fig: par fit Soleus}, in the plane $(f_0,r)$, we can see that the coloured lines and symbols (with error bars) obtained with the minimal model without fixing the size $N$, fall very close to the mean field hyperbolae with fixed $N$ (gray solid lines), and the (mean) estimates (gray dots).  
It can also be observed that the distributions of $N$ that falls into the physiological interval of accepted values, displays a clear peak around the values obtained previously with independent measurements. 

\subsection{Heterogeneity degree estimation}
\label{sec: Results het degree est}

In the study of the characterisation of the molecular aspects of actin mutations, we can analyse samples constituted of either a wild type or a mutant filament, and then investigate the heterofilament constituted of some mutant monomers and some healthy ones, to estimate the heterogeneity degree of the filament. 
In this Section we consider an heterogeneous ensemble of fixed unknown size $N$, constituted of two populations of motors, fast and slow, with an unknown proportion $p$ of fast motors, such that: $N=p N_F+(1-p)N_S$. 
We assume the possibility of independently analysing two homogeneous ensembles (of any size $N_j$, $j=\{F,S\}$) of both the populations of motors that constitute the mixed ensemble. 
Appling the procedure described in Section~\ref{sec: Results hom ensemble}, we provide an estimation of the population-dependent parameters, such as the single motor force, and the duty ratio, for each homogeneous ensemble.
With this knowledge we are able to estimate the model parameter $p$ that determines the degree of heterogeneity of the mixed state, as well as the size $N$ of the mixed ensemble.
To this aim, we have designed an estimation procedure that employs the theoretical probability distribution of the heterogeneous ensemble force computed in \eqref{eq: PF gauss}.

In order to validate our approach with synthetic data, we proceed as follows: 
\begin{enumerate}
    \item 
    We generate the synthetic data of the isometric force development for the wild-type (fast, $F$) and mutant (slow, $S$) populations (of fixed size). The mechano-kinetic parameters are chosen to mimic the characteristics of the actin-myosin interactions we aim to investigate.   
    \item 
    We characterise the mechano-kinetic performance (in terms of single motor force, ensemble duty ratio, and rate of transition through the attachment-detachment cycle) of the pure isoform ensembles, with the optimisation procedure described in Section~\ref{sec: Results hom ensemble}.
    Such a procedure is based on modelling the isometric force development using the stochastic model of Section~\ref{sec: masterEq}, and then on the inference of the mechano-kinetic parameter of the motors, for the fast and slow populations independently, processing their performance. 
    \item 
    We then generate a set of synthetic data of the force development of a heterogeneous ensemble constituted of two populations of motors, with fixed proportions each. Specifically we consider a fixed size $N=20$ of the heterogeneous ensemble, and different proportions $p$ of fast motors.
    \item 
    We analyse each data set to characterise the size and the degree of heterogeneity of the mixed ensemble, under the assumption that they are unknown. This analysis enables the validation of the estimation procedure on simulated data prior to its application to the experimental data sets.
    The analysis consists of an optimisation procedure, based on a simulated annealing algorithm \cite{cardoso1996simplex}, to estimate the best fit value of parameters $p$ and $N$.
\end{enumerate}

\begin{figure}[htbp]
    \centering
    \begin{subfigure}[b]{0.49\textwidth}
        \centering
        \includegraphics[width=\textwidth]{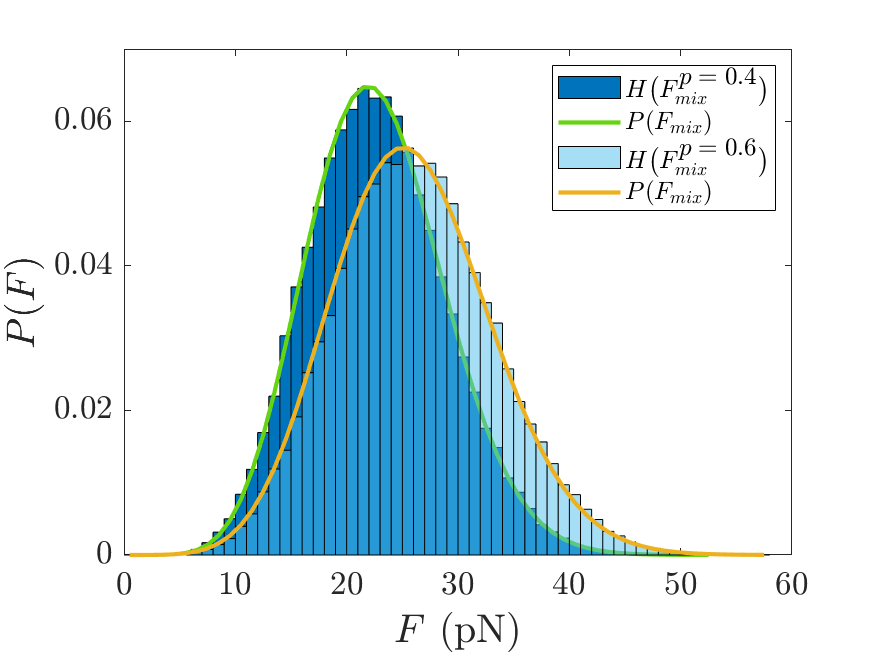}
        \caption{}
    \label{fig: fitData Mixed}
    \end{subfigure}
      \begin{subfigure}[b]{0.49\textwidth}
         \includegraphics[width=\textwidth]{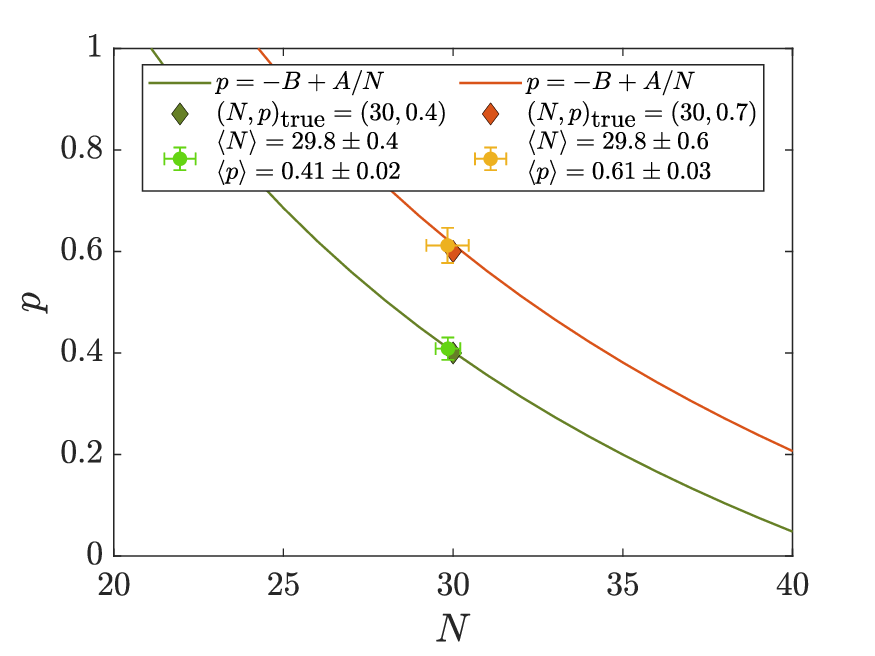}
        \caption{}
    \label{fig: pVsN Mixed}    
    \end{subfigure}
    \caption{
    \textbf{Heterogeneity degree estimation}\\
    We investigated two different cases: a mixed ensemble with a proportion of fast motors of either $p=0.4$ and $p=0.6$.
    \textbf{Panel} \textbf{(a)}: the histogram $H(F_{\textnormal{mix}})$ generated from the synthetic data of the force around the isometric plateau, for an heterogeneous ensemble of $N=30$ with $p=0.4$ (dark blue bars) and $p=0.6$ (light blue bars) fast motors. 
    The mechano-kinetics of the fast and slow populations are the same of the ones in figure~\ref{fig: ensemble force}. The green (yellow) curve is the best fit $P(F_{\textnormal{mix}})$ obtained with the optimisation procedure for the lower (higher) $p$ case.
    \textbf{Panel} \textbf{(b)}: the parameters resulting from the best fit evaluation are plotted in the plane $(N,p)$ for the two considered case. The green and yellow symbols are the mean values and SDs obtained as the average from multiple independent realisations of the optimisation procedure, for $p=0.4$ and $p=0.6$ respectively.
    The green and orange diamonds are the true parameter values for the two cases at hand. 
    The green and orange solid lines represent the mean field relation between $p$ and $N$.
    }
    \label{fig: fit het ensemble}
\end{figure}

Specifically, during the optimisation procedure, for each guess of the parameter $\tilde{p}$, we determine the corresponding value of the heterogeneous system size $\tilde{N}$ using Eq.~\eqref{eq: MF hyp}.
We consider that $N$ is an integer number, and it can take values only within a finite set (we select the set to be $N \in [8,32]$). 
In evaluating Eq.~\eqref{eq: MF hyp} we consider the fast and slow parameters $(f_0, r)_{\{F,S\}}$ obtained with the mentioned analysis on homogeneous ensembles.
Then, we compute the corresponding probability distribution $P(F)=P\big(F( \tilde{p}, \tilde{N}(\tilde{p}))\big)$ as a function of the model parameters, to be compared with the empirical histogram $H(F)$ of the force fluctuations around the isometric plateau. 
The best fit value is essentially the one that minimise the distance $ \mathcal{L}= \sum_{F}\vert H(F)- P(F) \vert^2$.
We refer the reader to the Appendix~\ref{sec: App optimisation proc} for more details on the optimisation algorithm. \\
In Figure~\ref{fig: fit het ensemble} the results of the heterogeneity degree estimation is shown, for two different mixed data sets, one with a lower proportion of the fast population ($p=0.4$) and one with a higher proportion ($p=0.6$). 
For these two conditions, each histogram $H(F_{\mathrm{mix}})$ (darker colour for $p=0.4$) is compared with the best fit $P(F_{\mathrm{mix}})$ (green curve for $p=0.4$, yellow for $p=0.6$). 
The green and orange diamonds in the figure are the true values of the parameters $p=0.4$ and $p=0.6$ respectively.
The procedure that selects a value for the proportion of fast motor $p$ and the system size $N$ is repeated multiple times, and the estimated values provided in Figure~\ref{fig: pVsN Mixed} (green symbol with error bars for $p=0.4$ and yellow symbol with error bars for $p=0.6$) are the mean values and SDs computed from multiple independent iterations of the stochastic optimisation algorithm.
The solid green and red lines represent the mean field relations between $p$ and $N$ as a function of the slow and fast populations parameters (the single motor force and the duty ratio), obtained with the estimation procedure on the homogeneous ensembles detailed in Section~\ref{sec: Results hom ensemble}.

\subsection{Inferring the performance of a pure isoform via the characterisation of the heterogeneous ensemble}
\label{sec: Results het fast}

\begin{figure}[htbp]
    \centering
    \begin{subfigure}[b]{0.49\textwidth}
        \centering
        \includegraphics[width=\textwidth]{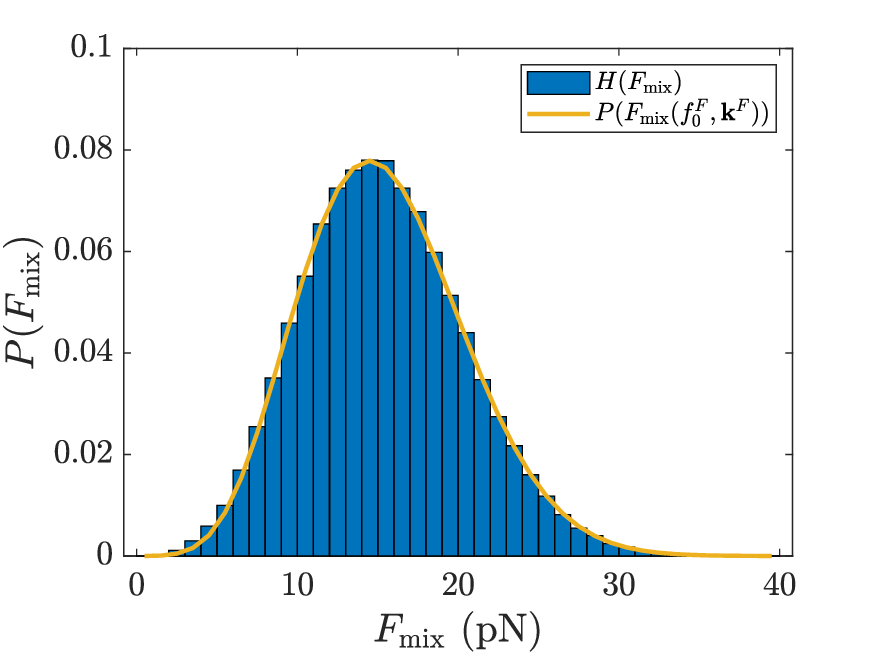}
        \caption{}
    \label{fig: fit data fast est p0.4}
    \end{subfigure}
    \caption{\textbf{Estimation of fast ensemble parameters from the analysis of the heterogeneous ensemble performance}\\
    The heterogeneous ensemble performance is investigated to determine the fast ensemble parameters. 
    The histogram $H(F_{\textnormal{mix}})$ generated from the synthetic data of the force around the isometric plateau is compared with the best fit $P(F_{\mathrm{mix}}(f_0^F, \vb{k}_F))$ obtained with the optimisation procedure of the fast ensemble mechano-kinetic parameters, under the assumption of a known system size $N=20$ and the heterogeneity degree $p=0.4$.
    }
\label{fig: fit mix fast est}
\end{figure}

In this section we consider a rather typical scenario of two different myosin isoforms that could be found in the experimental sample. 
We consider the case in which pure isoform ensemble experiments only with the slow population can be performed, but a sample of pure fast isoform ensemble is not easily achievable. 
We have designed a procedure to infer the mechano-kinetic performance of the motor ensemble that cannot be directly investigated (in this case the fast one), from the inspection of an heterogeneous ensemble of size $N$, with a fixed proportion $p$ of fast myosin isoform.
The theoretical expression of the probability distribution of the ensemble force fluctuations around the isometric plateau is still expressed by relation \eqref{eq: PF}, but in this case the fraction of fast motors $p$ can be set equal to the one known from the literature (for the specific case under investigation), and the slow population parameters can be estimated following the procedure illustrated in Section~\ref{sec: Results hom ensemble} for an homogeneous ensemble of size $N$. 
The free fitting parameters are the ones related to the fast populations, i.e. $(f_0, \vb{k})_F$, that are estimated with the same optimisation algorithm described in Section~\ref{sec: Results het degree est}. This algorithm minimises a loss function like \eqref{eq: loss fun}, which depends also on the system size $N$.
In Figure~\ref{fig: fit mix fast est} are shown the results of the procedure, and the estimated parameters are listed in Table~\ref{tab: fit mix fast est} .

\begin{table}[h!]
	\begin{center}
		$\begin{array}{cccccc}
			\toprule
			\text{} & f_0^F \ (\textnormal{pN}) &  r_F & \phi_F \ (\textnormal{s}^{-1}) & F_0 \ (\textnormal{pN}) 
			\\
			\midrule
			\text{True parameters}& 5 & 0.39 & 6.1 & 21.8 \\
			\text{(FAST ensemble)}& & & &\\
			& & & & \\
			\text{Estimated par. }& 5.05 \pm 0.09 & 0.40 \pm 0.01 & 6.0 \pm 0.3 & 21.8 
			\\
			\text{(FAST ensemble)}& & & &\\        
			\bottomrule
		\end{array} $
    \caption{\textbf{Estimated parameters via the optimisation procedure fed with simulated data, for a fast ensemble.}\\
    The parameters are: the force of a single motor $f_0^F$, the duty ratio $r_F$ and the rate of transition through the attachment–detachment cycle $\phi_F$ of the fast ensemble. Mean and SDs are computed from different independent realisations of the optimisation procedure. Errors are below $10^{-3}$ if not explicitly provided.
		}
		\label{tab: fit mix fast est}
	\end{center}
\end{table}

\section{Discussion}
\label{sec:discussion}

The studies reported in this paper concern isometric force generation in the sarcomere, where actin-myosin cyclic interactions take place within a highly organized assembly of regulatory and structural proteins. 
Some of the molecular aspects of this extremely articulated system can be experimentally investigated \textit{in vitro} thanks to a unique tool, the synthetic nanomachine, that involves a DLOT setup, a single bead-attached actin filament, and a support that carries the molecular motors. 
In this respect, the DLOT acts as a force transducer and the motors support acts as a displacement transducer.
This setup is able to reproduce, in well-controlled conditions, the collective mechanical behaviour of a small ensemble of myosin motors during isometric contraction, closely mimicking their function within the half-sarcomere, the functional unit of striated muscle. 
The nanomachine therefore represents an intermediate experimental scale between single-molecule assays and intact muscle investigation.
In this context, theoretical modelling is essential to allow for the inference of microscopic parameters from the analysis of meaningful mesoscopic quantity, as the ensemble force.

In this paper, we have addressed the fundamental aspect of heterogeneity of experimental samples.\\
In the model Section \ref{sec: Model} we computed the expression of the probability distribution $P(F)$ of the force exerted by molecular motors working collectively in a small ensemble of finite size $N$, in terms of the mechano-kinetic parameters of the system. The knowledge of $P(F)$ is crucial to characterise the statistics of the force fluctuations around the isometric plateau, which is the key quantity entering the estimation procedure we proposed for both homogeneous and heterogeneous ensembles.\\
In Section \ref{sec: Results hom ensemble} of the Results we have detailed the estimation procedure that we applied to synthetically generated data of actin-myosin interaction for a small homogeneous ensemble of unknown size, demonstrating how we are able to estimate both the mechano-kinetic parameters that underlie the stochastic force generation, and the system size $N$. This result is essential to the analysis of the heterogeneous ensemble. 
Furthermore, in Section \ref{sec: Results exp}, we also applied the procedure to experimental data of an ensemble of myosin motors purified from mammalian skeletal muscle, specifically an homogeneous ensemble of soleus HMMs (a slow myosin isoform)~\cite{buonfiglio2024force}. On these experimental data, we have showed that {\it (i)} the two-state model is suitable to describe the force fluctuations around the isometric plateau of an ensemble of motors working at room temperature condition, and {\it (ii)} the estimate of the system size is compatible with the experimental counts of the number of bond rupture events in rigor (ATP-free) conditions \cite{pertici2018myosin}.\\
In Section \ref{sec: Results het degree est} and in Section \ref{sec: Results het fast} we focused on the heterogeneous ensemble, considering two different possible scenarios (detailed below) where our results managed to characterise the molecular parameters of an heterogenous system. 
These two settings correspond to distinct inverse problems: estimating the composition of a mixed system when both homogeneous populations are characterised, and inferring the properties of an uncharacterised population when the mixture composition is known. \\
In Section \ref{sec: Results het degree est} we applied the estimation procedure to determine the heterogeneity degree of an ensemble in which fast and slow populations work collectively as independent force generators. 
This analysis is possible if we have experimental access to samples of homogeneous populations, which we can investigate independently to identify their mechano-kinetic performance through the fitting procedure discussed in Section \ref{sec: Results hom ensemble}. Mechano-kinetic performance are required to compute the theoretical probability distribution of the heterogeneous ensemble force, following Eq.~\eqref{eq: PF gauss}.\\
A possible relevant application of estimating the heterogeneity degree of an ensemble of motors is to investigate actin mutations. In \cite{pertici2026mechanics} the authors used the synthetic myosin-based nanomachine that mimics the loading conditions of $\gamma$-actin-myosin interactions \textit{in situ}, and performed comparative measurements on (i) wild-type $\gamma$-actin and (ii) $\gamma$-actin carrying the E334Q mutation associated with non-muscle actinopathies. 
Then, the authors characterised the mechano-kinetic parameters of the homogeneous wild-type and the mutant ensembles, where the mutant ensemble is composed of motors interacting with an actin filament with $100\%$ of mutant monomers.
A hybrid heterofilament ($50\%$ WT and $50\%$ mutant monomers) could be also studied, but in such a case the theoretical analysis that relies on the homogeneity of the mechano-kinetic features of the ensemble does not provide correct predictions.
Such a hybrid heterofilament can be generated when the F-actin polymerization is initiated with a fixed ratio of wild-type and mutant monomers. The fraction $p$ of mutant monomers within the heterogeneous filament reflects the initial ratio only if the mutation does not influence the polymerisation rate. 
For example cytoskeletal $\gamma$-actin filament reconstituted with a fraction $p=0.5$ of mutation p.E334Q of cytoskeletal $\gamma$-actin monomers shows a polymerisation rate 1.1-fold lower than the wild-type one~\cite{greve2024non}.
In the present paper, we have demonstrated how to address these experimental scenarios with a heterogeneous ensemble of motors of unknown size $N$, interacting with an actin filament composed of an unknown fraction $p$ of mutant monomers.\\
A different scenario is when the direct investigation of the two homogeneous ensembles is not experimentally achievable.
This is the case of HMMs purified from the majority of skeletal or cardiac muscle samples, that express a combination of different myosin isoforms.
We focused on the case when only the slow population ensemble can be directly investigated, inferring the mechano-kinetic parameters of the fast population from the analysis of the heterogeneous ensemble. The results of this procedure are presented in Section \ref{sec: Results het fast}. 
As an experimental reference, we can consider an array of myosin motors purified from bovine atrial cardiac muscle, which contains a certain proportion of the slow myosin isoform ($\beta$-MHC), in addition to a majority of the faster isoform ($\alpha$-MHC). 

In a future work we will apply the proposed methodology to experimental data, acquired with the synthetic nanomachine assembled with an heterogeneous ensemble of HMMs purified from mammalian cardiac muscle, with the aim to characterize the fast isoform mechano-kinetic parameters.

\appendix

\section{The Optimisation procedure: homogeneous ensemble}
\label{sec: App optimisation proc}

We here summarise the main steps of the optimisation procedure to estimate the ensemble size $N$ and its physiological mechano-kinetic parameters, like the force of a single motor $f_0$, the duty ratio $r$ and the rate $\phi$ through the attachment-detachment cycle.
The first step is to estimate the rate of force development $1/\tau$, that can also be computed from the inverse of the time rise $t_r$ from the $10\%$ to the $90\%$ of the isometric force plateau ($1/\tau= 2.2/t_r$). 
Then, we identify the portion of the force time series that represents the fluctuating dynamics around the stationary state, by removing the transient of the force development.
To characterise the stationary statistical features of the fluctuating ensemble force, we compare the histogram $H(F)$ with the theoretical expression of $P(F)$, Eq.\eqref{eq: P(Fj)}.
We perform an optimisation procedure based on nonlinear simplex and simulated annealing (SIMPSA) algorithm \cite{cardoso1996simplex}. 
The SIMPSA algorithm is designed for the global optimization of multi‑variate nonlinear constrained problems. 
Specifically, it attempts to solve problems of the form:
\begin{equation*}
    \min_{X} h(X) \ \text{subject to: } l_b \leq X \leq u_b  \ .
\end{equation*}
In the present case, the set of free fitting parameters $X=(f_0, k_+)$ are the force of a correctly oriented motor and the attachment kinetic rate constant.
The detachment kinetic rate constant $k_-$ can be computed by imposing the mean field rate of force development: $k_-= 1/\tau -k_+$.
We select an interval for each parameter, within a physiologically reasonable range: the lower bound is set to $l_b= (0.1, 0.1)$, while the upper bound is set to $u_b =(10, 50)$.
The algorithm is implemented with the default set options. \\
We iterate the following procedure. 
For each parameters guess we compute two quantities: the detachment kinetic rate constant $k_-=1/\tau -k_+$, and the system size $N$, which is bounded by the mean field relation at the asymptotic plateau (for $t \to \infty$) as in Eq.~\eqref{eq: N hom}.
We also impose the constraint that $N$ has to be an integer, since it represents the number of motors in the system, and that it falls into a physiological range set to be $[8,32]$.
Then, we calculate the probability distribution  $P\big(F(N, f_0,\vb{k})\big)$ of the force around the isometric plateau, as a function of the mentioned parameters with relation~\eqref{eq: P(Fj)}, and we evaluate the loss function computing the distance between the numerical distribution $H(F)$ and the theoretical one $P(F)$. The latter is given by the mean squared error:
\begin{equation}
    \label{eq: loss fun}
    \mathcal{L}= \frac{1}{L} \sum_{F} |H(F)-P(F)|^2     
\end{equation}
where $\vert \cdot \vert$ is the usual quadratic norm, $H(F)$ is obtained from the synthetic data of the stochastic force at the isometric plateau, and $P(F)$ is the theoretical counterpart given by Eq.\eqref{eq: P(Fj)}.
The procedure converges when a minimum of $\mathcal{L}$ is found in the chosen interval of parameters values $[l_b,u_b]$, and the corresponding parameters are the best fit parameters $(\hat{N}, \hat{f_0}, \hat{\vb{k}})$ selected by the procedure. 
The algorithm is stochastic, therefore even with the same starting point we obtain different results. We thus perform multiple $m$ independent realisations of the optimisation procedure, with different starting point inside the interval $[l_b,u_b]$, and express the best fit parameters $(N, f_0, \vb{k})$ as the mean $\pm$ SD of the results from independent realisations of the estimation procedure $\braket{N}=\frac{1}{m}\sum_{i=1}^m N_i$, $\braket{f_0}=\frac{1}{m}\sum_{i=1}^m f_{0,i}$, $\braket{\vb{k}}=\frac{1}{m}\sum_{i=1}^m \vb{k}_i$.
We accept the fit results if the loss function is lower than a certain tolerance, which is suitably chosen dependently on the numerical histogram binning (tolerance is set equal to $3 \times 10^{-7}$ for the slow ensemble and $3 \times 10^{-8}$ for the fast ensemble).

\section{Validation of the two-state model for actin-myosin interaction}
\label{sec: App min mod validation}

We propose a validation method for the minimal interaction scheme we adopt to model actin-myosin interactions. 
Let us thus consider a phenomenological three-state model for the isometric force generation, and we generate data sets of the ensemble force exerted by $N$ motors (we refer the reader to the Methods in \cite{buonfiglio2024force, buonfiglio2025resolving} for more details on the stochastic three-state model for actin-myosin interactions). 
Each of the $N$ motors undergoes the following kinetic scheme:
\begin{equation}
\label{eq: three state mod kinetic scheme}
\ce{D <=>[{k_1}][{k_{-1}}] A_1 <=>[{k_2}][{k_{-2}}] A_2 ->[{k_3}]  D}
\end{equation}
where the $k_j$ are the kinetic rate constants.
We here report the results of the mean field analysis of the population dynamics of the three-state model; specifically, the equilibrium concentration of the force-generating states is:
\begin{equation}
    \label{eq: three state mod}
    y^*= \frac{k_1}{k_1+G}\frac{k_{-2}+ k_3}{k_2+k_{-2}+k_3} \ ,\qquad z^*= \frac{k_1}{k_1+G}\frac{k_2}{k_2+k_{-2}+k_3}
\end{equation}
with $G=(k_{-1}(k_{-2}+k_3)+k_2 k_3)/(k_2+k_{-2}+k_3)$. 
Therefore, the duty ratio of the system is $r= k_1/(k_1+G)$. 
If we compare this expression with the duty ratio of the two-state model adopted in this work, $r=k_+/(k_+ + k_-)$, we get: 
\begin{equation}
\label{eq: two-state vs three state}
\begin{cases}
    k_+=k_1 \\
    k_-=G.
\end{cases}
\end{equation} 

\begin{figure}[t]
    \centering
    \begin{subfigure}[b]{0.29\textwidth}
        \centering
        \includegraphics[width=\textwidth]{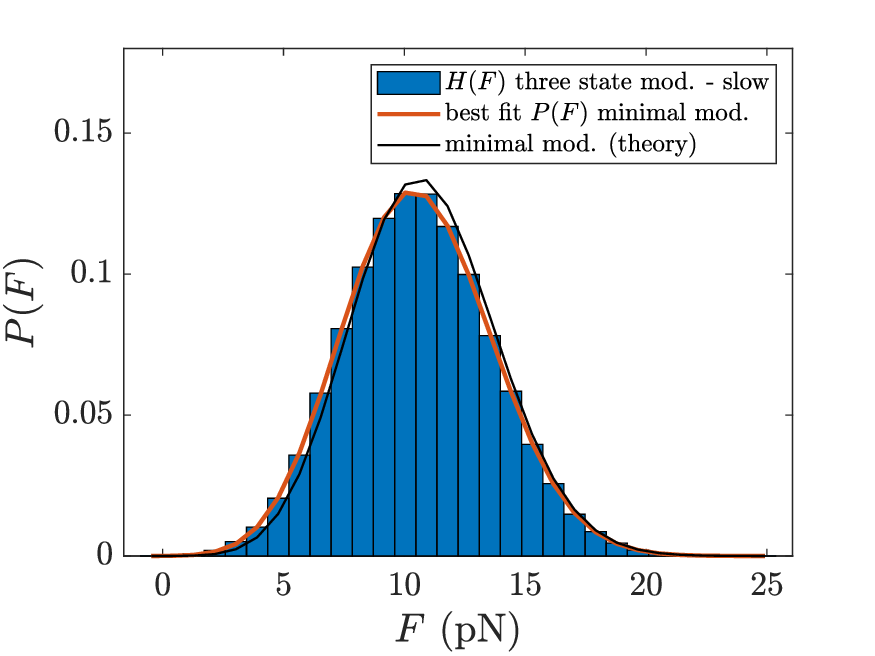}
        \caption{}
    \label{fig: fitData SYNTH-slow}
    \end{subfigure}
      \begin{subfigure}[b]{0.29\textwidth}
        \centering
        \includegraphics[width=\textwidth]{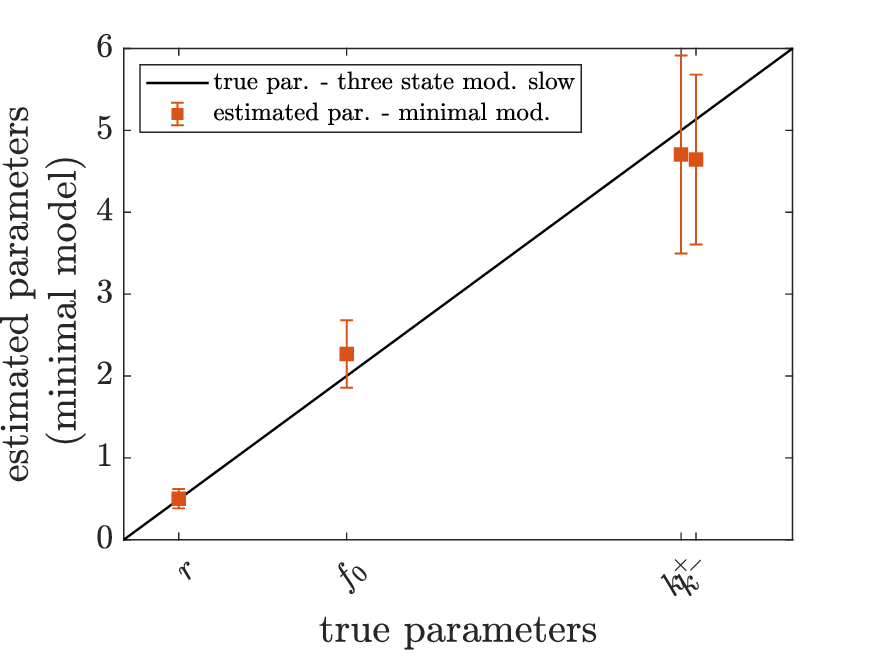}
        \caption{}
    \label{fig: }
    \end{subfigure}
      \begin{subfigure}[b]{0.29\textwidth}
         \includegraphics[width=\textwidth]{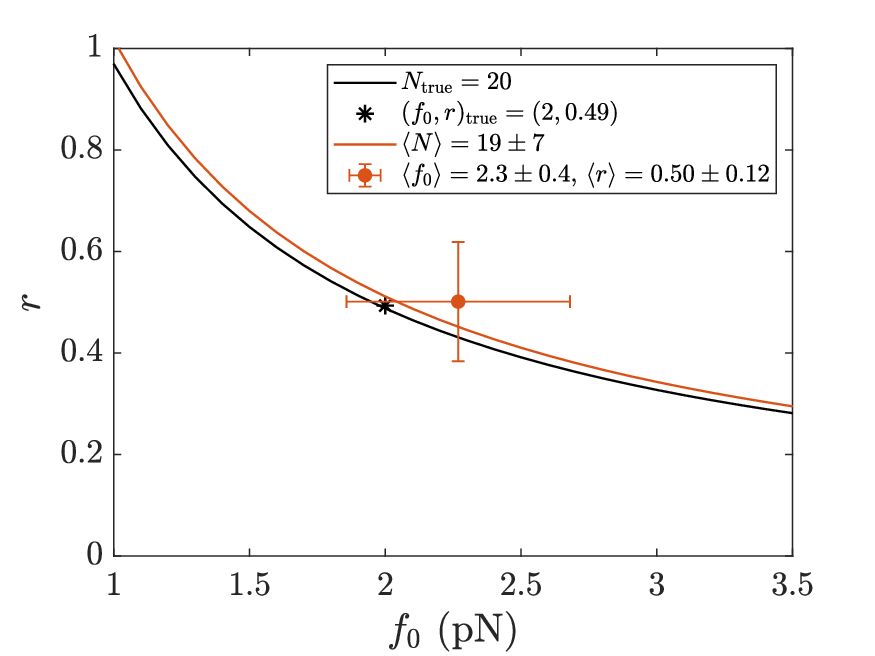}
        \caption{}
    \label{}    
    \end{subfigure}
     \begin{subfigure}[b]{0.29\textwidth}
        \centering
        \includegraphics[width=\textwidth]{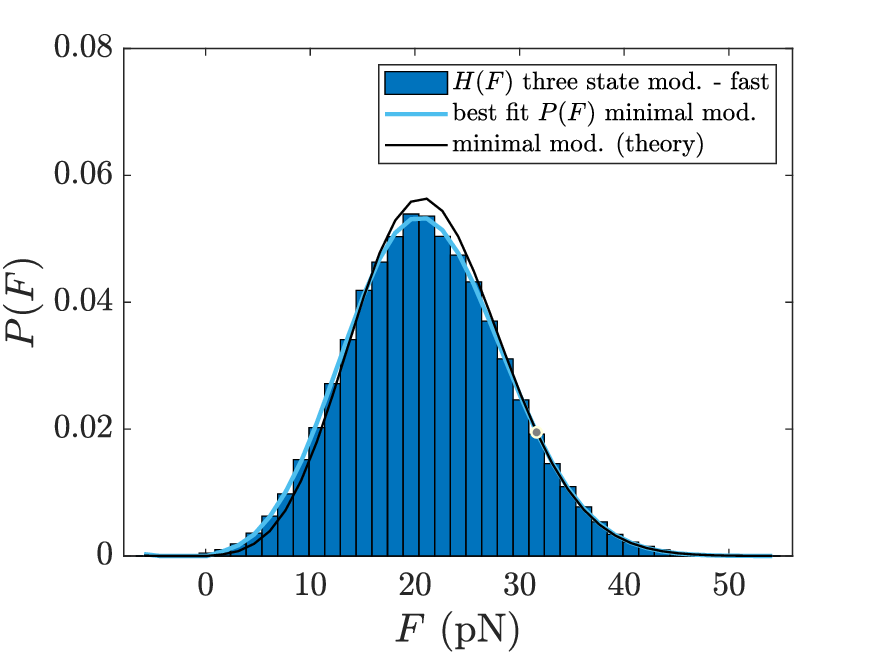}
        \caption{}
    \label{}
    \end{subfigure}
    \begin{subfigure}[b]{0.29\textwidth}
        \centering
        \includegraphics[width=\textwidth]{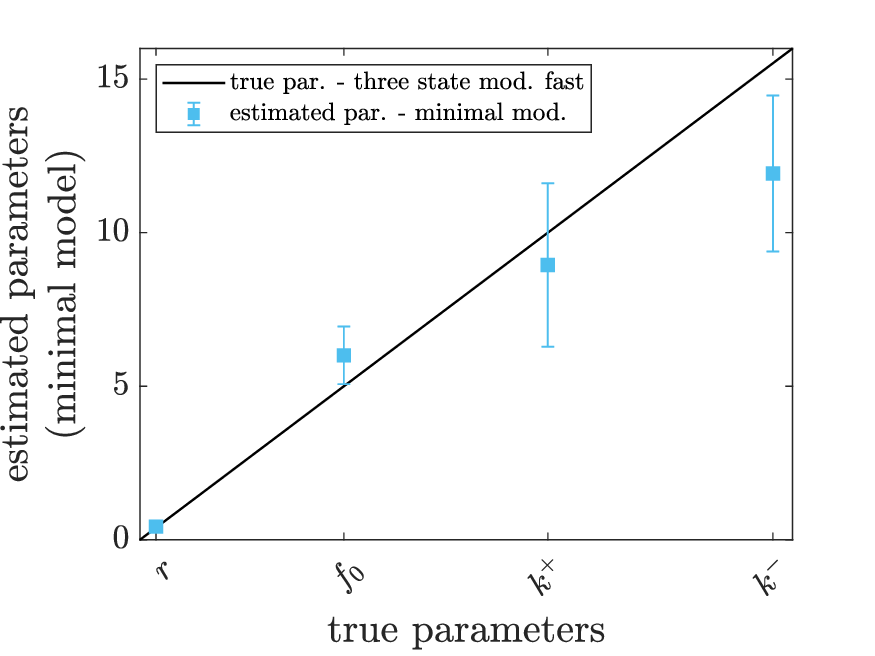}
        \caption{}
    \label{}
    \end{subfigure}
    \begin{subfigure}[b]{0.29\textwidth}
        \centering
        \includegraphics[width=\textwidth]{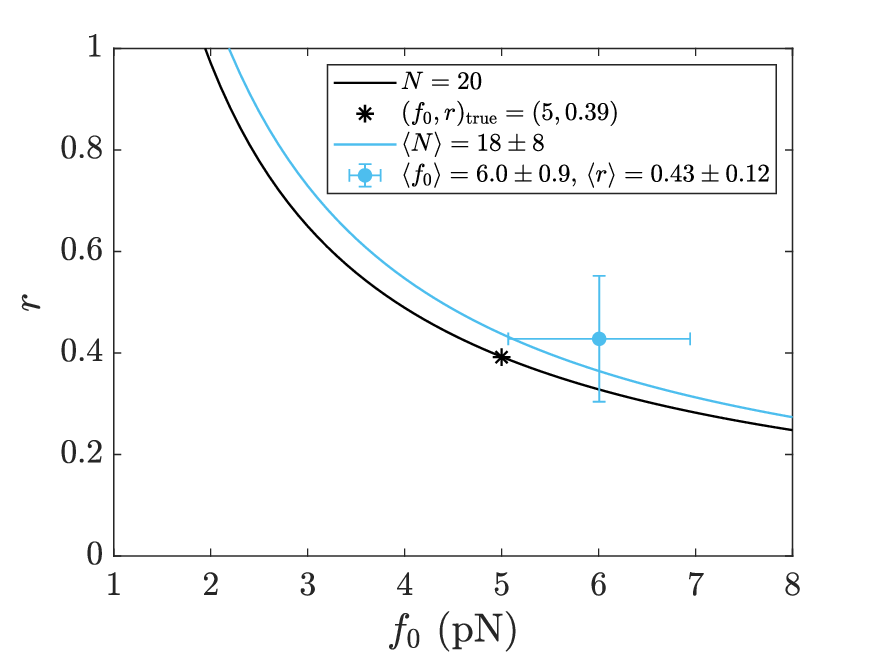}
        \caption{}
    \label{}
    \end{subfigure}
    \caption{
    \textbf{Fitting the three-state model data with the two-state model.}
    Synthetic data generated with a three-state model for cyclic actin-myosin interactions, fitted with the inverse scheme that employs the two-state model for homogeneous ensembles.
    \textbf{Panels} \textbf{(a)}, \textbf{(d)}: the histogram $H(F)$ generated from the three-state model data (blue bars) is compared with the theoretical best fit $P(F)$ for the two-state model (coloured curve), for a typical "slow" (light red) and "fast" (light blue) set of kinetic parameters $\vb{k}$.
    \textbf{Panels} \textbf{(b)}, \textbf{(e)}: the estimated mechano-kinetic parameters $(f_0, r, k_+,k_-)$ are plotted against their true value, for the slow (light red) and fast (in light blue) ensemble, respectively. 
    \textbf{Panels} \textbf{(c)}, \textbf{(f)}: in the plane $(f_0,r)$ we compare the true values of the parameters (black symbol) that lies on the hyperbolae with the true system size (black line), and the hyperbolae (coloured line) and symbol (coloured) resulting from the best fit, for the slow (light red) and fast (light blue) ensemble, respectively.
    }
    \label{fig: fit min mod validation}
\end{figure}

In Figure~\ref{fig: fit min mod validation} we present the outcome of the application of the optimisation procedure performed with the inverse scheme that employs the two-state model for homogeneous ensemble, to synthetic data generated with a three-state model. 
We consider two different kinetics, a slow ensemble (upper panels) and a fast ensemble (lower panels). For each ensemble we plotted the the histogram $H(F)$ computed from the stochastic simulations of the force time series against the best fit $P(F)$. We also plotted the estimated parameters of the two-state model ($f_0,r,k_+,k_{-}$) against the corresponding "true" value obtained from the relations \eqref{eq: two-state vs three state}, and the accord is reasonable.
Lastly, in the plane $(f_0,r)$ we compare the true values of the
parameters (black symbol) that lies on the hyperbolae computed with the true system size (black line), with the hyperbolae (coloured line) and symbol (coloured) resulting from the optimisation scheme. In the legend of the plot the numerical values of the parameters (true and fitted) are also reported.\\

\noindent
\textbf{Data Accessibility.}
All codes for this paper can be found at: \\ https://github.com/ValentinaBuonfiglio/Reverse-engineering-of-mechano-kinetic-parameters.\\

\noindent
\textbf{Funding statement.}
This work was supported by the European Union -  NextGeneration EU, within PRIN 2022, PNRR, Project No. P2022XPT32 “Regulation of striated muscle: a research bridging single molecule to organ (ReStriMus)”, CUP B53D23033290001.\\

\noindent
\textbf{Acknowledgments.}
We thank Prof. Vincenzo Lombardi for continuous discussion and insightful comments on the manuscript.\\

\noindent
\textbf{Competing interests.} The authors declare no competing interests.\\

\clearpage

\bibliographystyle{ieeetr}
\bibliography{Bibliography}

@article{huxley1957muscle,
  title={Muscle Structure and Theories of Contraction},
  author={Huxley, Andrew F},
  journal={Progress in Biophysics and Biophysical Chemistry},
  volume={7},
  pages={255--318},
  year={1957},
  publisher={Elsevier},
  url={https://doi.org/10.1016/S0096-4174(18)30128-8},
  doi={10.1016/S0096-4174(18)30128-8}
}

@article{huxley1971proposed,
  title={Proposed Mechanism of Force Generation in Striated Muscle},
  author={Huxley, Andrew F and Simmons, Ro M},
  journal={Nature},
  volume={233},
  number={5321},
  pages={533--538},
  year={1971},
  publisher={Nature Publishing Group},
  url={https://doi.org/10.1038/233533a0},
  doi={10.1038/233533a0}
}

@article{woledge1985energetic,
  title={Energetic Aspects of Muscle Contraction},
  author={Woledge, Roger C and Curtin, Nancy A and Homsher, Earl},
  journal={Monographs of the Physiological Society},
  volume={41},
  pages={1--357},
  year={1985}
}

@article{pertici2018myosin,
  title={A myosin II nanomachine mimicking the striated muscle},
  author={Pertici, Irene and Bongini, Lorenzo and Melli, Luca and Bianchi, Giulio and Salvi, Luca and Falorsi, Giulia and Squarci, Caterina and Boz{\'o}, Tam{\'a}s and Cojoc, Dan and Kellermayer, Mikl{\'o}s SZ and others},
  journal={Nature Communications},
  volume={9},
  number={1},
  pages={1--10},
  year={2018},
  publisher={Nature Publishing Group},
  url={https://doi.org/10.1038/s41467-018-06073-9},
  doi={10.1038/s41467-018-06073-9}
}

@article{pertici2021muscle,
  title={Muscle myosin performance measured with a synthetic nanomachine reveals a class-specific Ca2+-sensitivity of the frog myosin II isoform},
  author={Pertici, Irene and Bianchi, Giulio and Bongini, Lorenzo and Cojoc, Dan and Taft, Manuel H and Manstein, Dietmar J and Lombardi, Vincenzo and Bianco, Pasquale},
  journal={The Journal of Physiology},
  volume={599},
  number={6},
  pages={1815--1831},
  year={2021},
  publisher={Wiley Online Library},
  url={https://doi.org/10.1113/JP280976},
  doi={10.1113/JP280976}
}

@article{pertici2020myosin,
  title={A myosin II-based nanomachine devised for the study of Ca2+-dependent mechanisms of muscle regulation},
  author={Pertici, Irene and Bianchi, Giulio and Bongini, Lorenzo and Lombardi, Vincenzo and Bianco, Pasquale},
  journal={International Journal of Molecular Sciences},
  volume={21},
  number={19},
  pages={7372},
  year={2020},
  publisher={MDPI},
  url={https://doi.org/10.3390/ijms21197372},
  doi={10.3390/ijms21197372}
}

@article{schiaffino2011fiber,
  title={Fiber Types in Mammalian Skeletal Muscles},
  author={Schiaffino, Stefano and Reggiani, Carlo},
  journal={Physiological Reviews},
  volume={91},
  number={4},
  pages={1447--1531},
  year={2011},
  publisher={American Physiological Society Bethesda, MD},
  url={https://doi.org/10.1152/physrev.00031.2010},
  doi={10.1152/physrev.00031.2010}
}

@article{hall1927distribution,
  title={The distribution of means for samples of size n drawn from a population in which the variate takes values between 0 and 1, all such values being equally probable},
  author={Hall, Philip},
  journal={Biometrika},
  pages={240--245},
  year={1927},
  publisher={JSTOR},
  url={https://doi.org/10.2307/2331961},
  doi={10.2307/2331961}
}

@article{finer1994single,
  title={Single myosin molecule mechanics: piconewton forces and nanometre steps},
  author={Finer, Jeffrey T and Simmons, Robert M and Spudich, James A},
  journal={Nature},
  volume={368},
  number={6467},
  pages={113--119},
  year={1994},
  publisher={Nature Publishing Group UK London},
  url={https://doi.org/10.1038/368113a0},
  doi={10.1038/368113a0}
}

@article{ishijima1996multiple,
  title={Multiple-and single-molecule analysis of the actomyosin motor by nanometer-piconewton manipulation with a microneedle: unitary steps and forces},
  author={Ishijima, Akihiko and Kojima, Hiroaki and Higuchi, Hideo and Harada, Yoshie and Funatsu, Takashi and Yanagida, Toshio},
  journal={Biophysical Journal},
  volume={70},
  number={1},
  pages={383--400},
  year={1996},
  publisher={Elsevier},
  url={https://www.cell.com/biophysj/fulltext/S0006-3495(96)79582-6},
  doi={10.1016/S0006-3495(96)79582-6}
}

@article{gillespie1976general,
  title={A general method for numerically simulating the stochastic time evolution of coupled chemical reactions},
  author={Gillespie, Daniel T},
  journal={Journal of Computational Physics},
  volume={22},
  number={4},
  pages={403--434},
  year={1976},
  publisher={Elsevier},
  url={https://www.sciencedirect.com/science/article/pii/0021999176900413},
  doi={10.1016/0021-9991(76)90041-3}
}

@article{gillespie1977exact,
  title={Exact stochastic simulation of coupled chemical reactions},
  author={Gillespie, Daniel T},
  journal={The Journal of Physical Chemistry},
  volume={81},
  number={25},
  pages={2340--2361},
  year={1977},
  publisher={ACS Publications},
  url={https://doi.org/10.1021/j100540a008},
  doi={10.1021/j100540a008}
}

@article{buonfiglio2024force,
  title={Force and kinetics of fast and slow muscle myosin determined with a synthetic sarcomere--like nanomachine},
  author={Buonfiglio, Valentina and Pertici, Irene and Marcello, Matteo and Morotti, Ilaria and Caremani, Marco and Reconditi, Massimo and Linari, Marco and Fanelli, Duccio and Lombardi, Vincenzo and Bianco, Pasquale},
  journal={Communications Biology},
  volume={7},
  number={1},
  pages={361},
  year={2024},
  publisher={Nature Publishing Group UK London},
  url={https://doi.org/10.1038/s42003-024-06033-8},
  doi={10.1038/s42003-024-06033-8}
}

@article{cardoso1996simplex,
  title={The simplex-simulated annealing approach to continuous non-linear optimization},
  author={Cardoso, Margarida F and Salcedo, Romualdo L and De Azevedo, S Feyo},
  journal={Computers \& chemical engineering},
  volume={20},
  number={9},
  pages={1065--1080},
  year={1996},
  publisher={Elsevier},
  url={https://www.sciencedirect.com/science/article/pii/0098135495002219},
  doi={10.1016/0098-1354(95)00221-9}
}

@article{buonfiglio2025resolving,
  title={Resolving the kinetics of an ensemble of muscle myosin motors via a temperature-dependent fitting procedure},
  author={Buonfiglio, Valentina and Zagli, Niccol{\`o} and Pertici, Irene and Lombardi, Vincenzo and Bianco, Pasquale and Fanelli, Duccio},
  journal={Journal of the Royal Society Interface},
  volume={22},
  number={225},
  pages={2025--0040},
  year={2025},
  publisher={The Royal Society},
  url={https://doi.org/10.1098/rsif.2025.0040},
  doi={10.1098/rsif.2025.0040}
}

@article{greve2024non,
  title={The non-muscle actinopathy-associated mutation E334Q in cytoskeletal $\gamma$-actin perturbs interaction of actin filaments with myosin and ADF/cofilin family proteins},
  author={Greve, Johannes N and Marquardt, Anja and Heiringhoff, Robin and Reindl, Theresia and Thiel, Claudia and Di Donato, Nataliya and Taft, Manuel H and Manstein, Dietmar J},
  journal={Elife},
  volume={12},
  pages={RP93013},
  year={2024},
  publisher={eLife Sciences Publications Limited},
  url={https://doi.org/10.7554/eLife.93013.3},
  doi={10.7554/eLife.93013}
}

@article{pertici2026mechanics,
  title={Mechanics of blunting of actin--myosin interaction dynamics by the actinopathy-causing mutation E334Q in cytoskeletal $\gamma$-actin},
  author={Pertici, Irene and Buonfiglio, Valentina and Greve, Johannes N and Battirossi, Elena and Fanelli, Duccio and Manstein, Dietmar J and Bianco, Pasquale},
  journal={The Journal of Physiology},
  year={2026},
  publisher={Wiley Online Library},
  url={https://doi.org/10.1113/JP289622},
  doi={10.1113/JP289622}
}

@article{di2016update,
  title={{Update on the ACTG1-associated Baraitser--Winter cerebrofrontofacial syndrome}},
  author={Di Donato, Nataliya and Kuechler, Alma and Vergano, Samantha and Heinritz, Wolfram and Bodurtha, Joann and Merchant, Sabiha R and Breningstall, Galen and Ladda, Roger and Sell, Susan and Altm{\"u}ller, Janine and others},
  journal={American journal of medical genetics Part A},
  volume={170},
  number={10},
  pages={2644--2651},
  year={2016},
  publisher={Wiley Online Library},
  url={https://doi.org/10.1002/ajmg.a.37771},
  doi={10.1002/ajmg.a.37771}
}

@article{gorza1982myosin, 
    author = {Gorza, L and Sartore, S and Schiaffino, S},
    title = {Myosin types and fiber types in cardiac muscle. II. Atrial myocardium.},
    journal = {Journal of Cell Biology},
    volume = {95},
    number = {3},
    pages = {838-845},
    year = {1982},
    month = {12},
    issn = {0021-9525},
    doi = {10.1083/jcb.95.3.838},
    url = {https://doi.org/10.1083/jcb.95.3.838},
    eprint = {https://rupress.org/jcb/article-pdf/95/3/838/1641005/838.pdf},
}

@article{SYROVY1989441,
title = {Expression of myosin in atrial areas of the bovine myocardium},
journal = {Comparative Biochemistry and Physiology Part A: Physiology},
volume = {92},
number = {3},
pages = {441-443},
year = {1989},
issn = {0300-9629},
doi = {https://doi.org/10.1016/0300-9629(89)90589-6},
url = {https://www.sciencedirect.com/science/article/pii/0300962989905896},
author = {I. Syrový}
}

@article{reconditi2003conformation,
  title={The conformation of myosin head domains in rigor muscle determined by X-ray interference},
  author={Reconditi, Massimo and Koubassova, N and Linari, Marco and Dobbie, I and Narayanan, T and Diat, O and Piazzesi, Gabriella and Lombardi, Vincenzo and Irving, M},
  journal={Biophysical journal},
  volume={85},
  number={2},
  pages={1098--1110},
  year={2003},
  publisher={Elsevier}
}

@article{cummins1986myosin, 
author = {P Cummins  and S J Lambert },
title = {Myosin transitions in the bovine and human heart. A developmental and anatomical study of heavy and light chain subunits in the atrium and ventricle.},
journal = {Circulation Research},
volume = {58},
number = {6},
pages = {846-858},
year = {1986},
doi = {10.1161/01.RES.58.6.846},
URL = {https://www.ahajournals.org/doi/abs/10.1161/01.RES.58.6.846},
eprint = {https://www.ahajournals.org/doi/pdf/10.1161/01.RES.58.6.846}}

@article{medler2019mixing,
  title={Mixing it up: the biological significance of hybrid skeletal muscle fibers},
  author={Medler, Scott},
  journal={Journal of Experimental Biology},
  volume={222},
  number={23},
  pages={jeb200832},
  year={2019},
  publisher={The Company of Biologists Ltd},
  doi={10.1242/jeb.200832},
  ulr={https://doi.org/10.1242/jeb.200832}
}

@article{BEACH20141160,
title = {Nonmuscle Myosin II Isoforms Coassemble in Living Cells},
journal = {Current Biology},
volume = {24},
number = {10},
pages = {1160-1166},
year = {2014},
issn = {0960-9822},
doi = {https://doi.org/10.1016/j.cub.2014.03.071},
url = {https://www.sciencedirect.com/science/article/pii/S0960982214003935},
author = {Jordan R. Beach and Lin Shao and Kirsten Remmert and Dong Li and Eric Betzig and John A. Hammer}
}

\clearpage
\markboth{}{}

\end{document}